\documentclass{aa}  

\usepackage{graphicx}
\usepackage{txfonts}
\usepackage{bm}
\usepackage[colorlinks=true,allcolors=blue]{hyperref}
\newcommand{\figref}[1]{\figurename~\ref{#1}}

\begin{document}

   \title{Collapse of spherical overdensities in superfluid models of dark matter}

   %\subtitle{}

   \author{S. T. H. Hartman
   %\inst{1}
          \and
          H. A. Winther
          %\inst{1}
          \and
          D. F. Mota
          %\inst{1}
          }

   \institute{Institute of Theoretical Astrophysics, University of Oslo,
              PO Box 1029, Blindern 0315, Oslo, Norway
             % \\\email{stian.hartman@astro.uio.no}}
}
   \date{Received ; accepted }

% \abstract{}{}{}{}{} 
% 5 {} token are mandatory
 
  \abstract
   {}
   {We intend to understand cosmological structure formation within the framework of superfluid models of dark matter with finite temperatures. Of particular interest is the evolution of small-scale structures where the pressure and superfluid properties of the dark matter fluid are prominent. We compare the growth of structures in these models with the standard cold dark matter paradigm and non-superfluid dark matter.}
   {The equations for superfluid hydrodynamics were computed numerically in an expanding $\Lambda$CDM background with spherical symmetry; the effect of various superfluid fractions, temperatures, interactions, and masses on the collapse of structures was taken into consideration. We derived the linear perturbation of the superfluid equations, giving further insights into the dynamics of the superfluid collapse.}
   {We found that while a conventional dark matter fluid with self-interactions and finite temperatures experiences a suppression in the growth of structures on smaller scales, as expected due to the presence of pressure terms, a superfluid can collapse much more efficiently than was naively expected due to its ability to suppress the growth of entropy perturbations and thus gradients in the thermal pressure. We also found that the cores of the dark matter halos initially become more superfluid during the collapse, but eventually reach a point where the superfluid fraction falls sharply. The formation of superfluid dark matter halos surrounded by a normal fluid dark matter background is therefore disfavored by the present work.}
   {}

   \keywords{cosmology: theory - dark matter - large-scale structure of Universe}

   \maketitle
%
%-------------------------------------------------------------------

\section{Introduction}

   A universe with cold dark matter (CDM), a cosmological constant ($\Lambda$), and inflationary initial conditions forms the foundation of the standard $\Lambda$CDM paradigm that has proven successful at explaining a wide range of observables, such as the expansion history of the universe, the cosmic microwave background, formation of large-scale structure, the matter power spectra, and the abundance of light elements \citep{Tegmark2004, Planck2015, Cyburt2016}. Nonetheless, it is a phenomenological model that is ignorant of the origin of the cosmological constant and the identity of dark matter (DM), which remain two of the greatest mysteries in fundamental physics today.
   
   A number of challenges to $\Lambda$CDM have emerged as both observations and numerical simulations become increasingly more precise, especially on small scales. The cores of DM halos predicted from {\it N}-body simulations are denser and more cuspy than observed, and the number of dwarf galaxies in the Local Group is far smaller than expected from pure $\Lambda$CDM simulations. These issues are known as the too-big-to-fail, cusp-core, and missing satellite problems (see e.g., \citet{DelPopolo2017} and \citet{Bullock2017} and references therein). Another puzzling phenomenology on the scale of galaxies is the empirical baryonic Tully-Fisher relation (BTFR) \citep{McGaugh2000, McGaugh2005, Lelli2015}. This relates the baryonic mass of galaxies $M_b$ with the asymptotic circular velocity $v_c$ through $M_b \sim v_c^4$ and holds for many orders of magnitude with remarkably small scatter. The $\Lambda$CDM prediction for this relation is $M\sim v_c^3$ with the total mass $M$ from both baryons and DM \citep{McGaugh2012}. It is the latter that dominates the gravitational pull in galaxies, which only adds to the strangeness of the BTFR.
   
   Solutions to these problems within the framework of $\Lambda$CDM have been proposed by including baryonic physics \citep{Santos-Santos2015, Sales2016, Zhu2016, Sawala2016}, but it is unclear if they can completely cure the ails of $\Lambda$CDM. These processes are not yet fully understood and are difficult to model in simulations of galaxy formation, and their stochastic nature makes it even more puzzling as to how they can be responsible for the tight correlation in scaling relations, such as the BTFR.
   
   An alternative possibility is that the mismatch between observations and simulations is an indication of physics beyond the standard model, either through modified theories of gravity, the particle nature of DM, or both. An example of such a model is modified Newtonian dynamics (MOND) \citep{Milgrom1983b, Milgrom1983c, Milgrom1983a, Famaey2012}, in which the Newtonian law of gravity in low-acceleration regions is modified to explain the rotation curves of galaxies without the need of resorting to DM. One of its most appealing features is that the BTFR and its small scatter is a direct consequence of it. However, MOND and its relativistic extensions face challenges of their own on extragalactic scales where the CDM paradigm is successful \citep{Zuntz2010, Dodelson2011, Angus2013, Angus2014}. This has, somewhat ironically, motivated extended models of DM where MOND is an emergent fifth force on small scales \citep{Berezhiani2015, Khoury2016}. This is achieved by DM undergoing Bose-Einstein condensation on galactic scales and adding a coupling designed to give a MONDian long-range force between baryons mediated by phonons in the superfluid cores of galaxies. Outside galaxies, the DM fluid ceases to be superfluid, and the extra force disappears, preserving the success of CDM on large scales.
   
   Superfluid dark matter (SFDM) models are also interesting on more general grounds. From condensed matter physics, we know that self-interacting boson gases can become superfluid given sufficiently high densities and low temperatures. In the weakly interacting Bose gas, the critical temperature that marks the onset of superfluidity depends almost solely on the particle mass and number density.  We can therefore expect boson DM candidates with self-interactions to exhibit superfluid behavior in certain mass ranges.
   
   Observations of the large-scale structure of the universe strongly favor cold and collisionless DM, but for SFDM this is no longer the case since the transition in and out of the superfluid phase requires both self-interactions and finite temperatures. We must therefore be wary of how structure forms in SFDM. Studies of other DM models with pressure-like terms, such as fuzzy dark matter \citep{Hu2000, Schive2014, Schwabe2016, Mocz2017} and self-interacting dark matter \citep{Spergel2000, Elbert2015, Tulin2018}, find they can help remove the surplus of small-scale structure in $\Lambda$CDM. So far, there has been little work done on structure formation in SFDM and how it differs from conventional DM fluids. In this paper, we aim to provide preliminary answers to these questions by considering the spherical collapse of SFDM.
   
   The paper is organized as follows: in Section \ref{sec:method}, the equations for superfluid hydrodynamics used to describe the collapse of SFDM are introduced, as well as the critical temperature and the critical velocity, which are important for the superfluid phenomenology. The linear expansion of the superfluid equations was derived to better understand how superfluidity changes the behavior of the DM fluid. In Section \ref{sec:results}, the results are presented and discussed, and we draw our conclusions in Section \ref{sec:conclusions}.

%--------------------------------------------------------------------
\section{Method}
\label{sec:method}

\subsection{Superfluid hydrodynamics}
To describe a finite-temperature superfluid, we employed the superfluid hydrodynamic equations \citep{Taylor2005, Chapman2014}, which in proper coordinates and physical variables are;

\begin{equation}
\label{eq:SF_mass_conservation}
    \frac{\partial \rho}{\partial t} + \bm{\nabla}\cdot\bm{j} = 0,
\end{equation}

\begin{equation}
\label{eq:SF_entropy_conservation}
    \frac{\partial S}{\partial t} + \bm{\nabla}\cdot(S \bm{u}_n) = 0,
\end{equation}

\begin{equation}
\label{eq:SF_velocity}
    \frac{\partial \bm{u}_s}{\partial t} + \bm{\nabla}(\mu + \frac{1}{2}\bm{u}^2_s) = -\bm{\nabla}\Phi,
\end{equation}

\begin{equation}
\label{eq:SF_momentum_conservation}
\begin{split}
    \frac{\partial \bm{j}}{\partial t} &+ \bm{\nabla}P + \rho_s(\bm{u}_s\cdot\bm{\nabla})\bm{u}_s + \rho_n(\bm{u}_n\cdot\bm{\nabla})\bm{u}_n \\
    &+ \bm{u}_s[\bm{\nabla}\cdot(\rho_s\bm{u}_s)] + \bm{u}_n[\bm{\nabla}\cdot(\rho_n\bm{u}_n)] = -\rho\bm{\nabla}\Phi,
\end{split}
\end{equation}

\begin{equation}
\label{eq:SF_energy_conservation}
\begin{split}
    \frac{\partial E}{\partial t} &+ \bm{\nabla}\cdot\Big[(U + \frac{1}{2}\rho_n u_n^2 + P)\bm{u}_n + \frac{1}{2}\rho_s u_s^2 \bm{u}_s \\
    & + \mu \rho_s(\bm{u}_s - \bm{u}_n)\Big] = -\bm{j}\cdot\bm{\nabla}\Phi.
\end{split}
\end{equation}
This set of equations describes the evolution of the fluid mass density $\rho$, entropy density $S$, superfluid velocity $\bm{u}_s$, momentum density $\bm{j}$, and energy density $E$ under the influence of the gravitational potential $\Phi$ sourced by matter and a cosmological constant,
\begin{equation}
    \nabla^2 \Phi = 4\pi G (\rho - 2\rho_{\Lambda}).
\end{equation}
Equations \eqref{eq:SF_entropy_conservation} and \eqref{eq:SF_energy_conservation} are degenerate in our set of equations if the solution is free of shocks, otherwise entropy is generated. The former is used in this work, but both are given for completeness.

A superfluid differs from a classical fluid in that it consists of two fluid components; the "superfluid" with density $\rho_s$ and velocity $\bm{u}_s$, and the "normal fluid" with density $\rho_n$ and velocity $\bm{u}_n$. The sum of the two component densities gives the total fluid density $\rho=\rho_n+\rho_s$, and likewise for momentum, $\bm{j}=\rho_n\bm{u}_n + \rho_s\bm{u}_s$. However, only the normal fluid transports entropy and thermal energy, as can be seen from Eqs. $\eqref{eq:SF_entropy_conservation}$ and $\eqref{eq:SF_energy_conservation}$, and the superfluid velocity evolves according to its own potential given in Eq. \eqref{eq:SF_velocity}, where the chemical potential is $\mu = [P+U-ST - \frac{1}{2}\rho_{n}(\bm{u}_s-\bm{u}_n)^2]/\rho$. Since there are two fluid components with separate velocity fields a superfluid can have two sound modes. One is called first sound and is associated with density perturbations, which we are familiar with from regular hydrodynamics. The other is called second sound and is associated with temperature perturbations. This is made possible by the fact that only the normal component carries entropy, hence the normal and superfluid components can oscillate in such a way that perturbations in temperature, and not density, are propagated through the fluid. As we will see it is this property that is responsible for difference in collapse of superfluid and non-superfluid DM.

The remaining variables in the above set of equations are pressure $P$, internal energy density $U$, and temperature $T$. In the limit $\rho_s = 0,$ they reduce to the Euler equations of fluid dynamics.

\subsection{Critical temperature and velocity, and equation of state}
When a boson gas is cooled below a critical temperature $T_c,$ the particles begin accumulating in the quantum ground state of the system and form a Bose-Einstein condensate (BEC). In the three-dimensional homogeneous and ideal Bose gas this critical temperature is
\begin{equation}
    T_c = \frac{2\pi\hbar^2}{m^{5/3} k_{\text{B}}}\left(\frac{\rho}{\zeta(3/2)}\right)^{2/3},
\end{equation}
where $\zeta(x)$ is the Riemann Zeta-function. This result holds approximately for weakly interacting gases as well \citep{Sharma2019}, apart from a small interaction-dependent shift \citep{Andersen2004} that we neglect.

The formation of a BEC does not automatically imply a superfluid. A further criterion must be satisfied as realized by \citet{Landau1941}. He assumed that if dissipation and heating happens through the creation of elementary excitations in the fluid, and if these excitations can no longer spontaneously appear the fluid will become superfluid. This gives the so-called Landau criterion and requires the relative motion $\bm{w}=\bm{u}_s-\bm{u}_n$ to be smaller than the critical velocity $v_c$,
\begin{equation}
    w < v_c = \min_{\bm{p}} \frac{\epsilon(\bm{p})}{p},
\end{equation}
where $\epsilon(\bm{p})$ is the energy of an elementary excitation with momentum $\bm{p}$. Clearly, we must have $v_c > 0$, otherwise any motion will destroy the superflow, and it no longer makes sense to refer to it as a superfluid. An ideal Bose gas can therefore not be superfluid since the elementary excitations are $\epsilon(\bm{p}) = p^2/2m$ so that $v_c=0$. In an interacting Bose gas, on the other hand, the condensation of the gas makes the energy spectrum phonon-like at small momenta, $\epsilon(\bm{p}) = c_s p$. The critical velocity is in this case finite, $v_c = c_s$, and we get superfluidity.

As $w$ approaches and exceeds the critical velocity, the superfluid flow begins to decay. This happens because a tangle of superfluid vortices, so-called quantum turbulence, forms and interacts with the excitations that make up the normal fluid, resulting in a dissipative mutual friction between the normal and superfluid components \citep{Skrbek2011,Skrbek2012,Barenghi2014}. This effect is not included in the equations for superfluid hydrodynamics and must be added through additional terms. However, this would require us to assume the dependence of this force on the fluid variables and specify the extra parameters introduced to our model (for examples of this in numerical studies of superfluid helium, see \citet{Doi2008}, \citet{Darve2012}, and \citet{Soulaine2017}). To capture the basic consequence of Landau's criterion relevant for this work, which is that the counterflow $w$ is limited by the critical velocity, we instead assume the mutual friction only takes place once the critical velocity is exceeded, and that the complicated processes taking place happen on time and length scales much shorter than we are considering. The mutual friction is therefore effectively instantaneous, and since it is dissipative, there is a conversion of kinetic energy into internal energy, heating the fluid and generating entropy. Stated more precisely, we enforce the superfluid critical velocity at every position in our numerical scheme by converting kinetic energy of the two fluid components (while conserving the total momentum) into internal energy and generated entropy so that $w < v_c$ is always satisfied. See Appendix \ref{app:numerical_enforce_vc} for further details.

\begin{figure}%[H]
    \centering
    \includegraphics[width=1\linewidth]{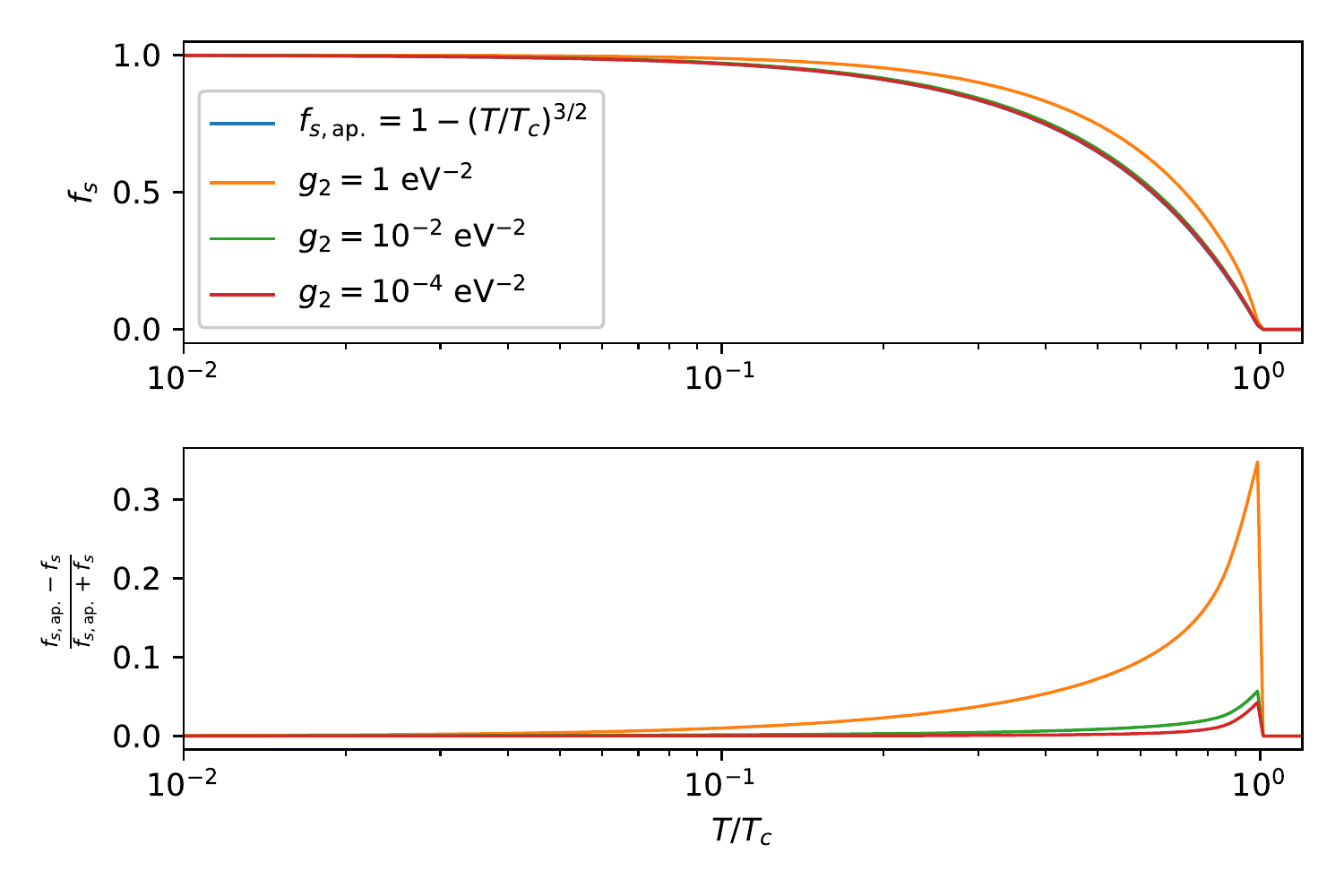}
    \caption{Superfluid fraction $f_s$ for the two-body interacting Bose gas calculated by \citet{Sharma2019} with $\rho=10^{8}\,\rho_{c0}$ and $m=1\,\text{eV}$ compared to the approximation $f_{s,\text{ap.}}=1-(T/T_c)^{3/2}$. For sufficiently weak interactions, $f_s$ can be approximated by the condensate fraction in an ideal Bose gas.}
    \label{fig:fs_exact_vs_approx}
\end{figure}

We must also specify the equation of state (EOS) that defines how the thermodynamic quantities depend on the temperature and particle density. In superfluids, the EOS is also a function of the counterflow $w$ \citep{Landau1987, Khalatnikov2000}, but we neglected this dependence and used the EOS corresponding to the $w=0$ limit. While this work is motivated by the superfluid DM model presented by \citet{Berezhiani2015}, it lacks a complete EOS at finite temperatures. We therefore employed the weakly interacting Bose gas with effective repulsive two- and three-body contact interactions as described by \citet{Sharma2019}, where the three-body case corresponds most closely to the model by \citet{Berezhiani2015}. Effective contact interactions can describe the s-wave scattering limit of more complicated interactions through the Born approximation, which makes this class of models a more general description of superfluids \citep{Pethick2008}. The coupling term between the DM fluid and baryons that gives rise to the emergent fifth force is not included in this work.
For computational speed, we approximated the EOS in the sub-$T_c$ regime by an ideal Bose gas with contributions from interactions at zero temperature. Notably, the superfluid fraction is approximated as the fraction of condensed particles in an ideal BEC, $f_s=\rho_s/\rho=1-(T/T_c)^{3/2}$. This might appear paradoxical since we already stated that an ideal Bose gas cannot be superfluid, but in the weakly interacting gas these quantities can be seen from \figref{fig:fs_exact_vs_approx} to be closely related. For strong interactions, the superfluid fraction can approach unity while the condensate fraction remains small, which is the case in superfluid $^4$He \citep{Glyde2013}, but this scenario is outside the scope of this paper. See Appendix \ref{app:eos_details} for further details on the EOS.

\subsection{Super-comoving variables}
Since we are interested in the evolution of the superfluid in an expanding space, we introduce the peculiar velocity $\bm{v}=\bm{u}-H\bm{r}$ and super-comoving variables \citep{Shapiro1998}, denoted by a tilde-sign, to rewrite the hydrodynamic equations in a more convenient form\footnote{The temperature and entropy in super-comoving variables are not given in \citet{Shapiro1998} (MS). We define them here as $\Tilde{T}=a^2T/T_*$ and $\Tilde{S}=a^3S/S_*$, where $T_*$ is a free parameter, $S_* = \rho_* v^2_*/T_*$, with $\rho_*$ and $v_*$ given in MS.}:

\begin{equation}
\label{eq:SF_mass_conservation_sc}
    \frac{\partial \Tilde{\rho}}{\partial \Tilde{t}} + \Tilde{\bm{\nabla}}\cdot\Tilde{\bm{j}} = 0,
\end{equation}

\begin{equation}
\label{eq:SF_entropy_conservation_sc}
    \frac{\partial \Tilde{S}}{\partial \Tilde{t}} + \Tilde{\bm{\nabla}}\cdot(\Tilde{S} \Tilde{\bm{v}}_n) = 0,
\end{equation}

\begin{equation}
\label{eq:SF_velocity_sc}
    \frac{\partial \Tilde{\bm{v}}_s}{\partial \Tilde{t}} + \Tilde{\bm{\nabla}}(\Tilde{\mu} + \frac{1}{2}\Tilde{\bm{v}}_s^2) = -\Tilde{\bm{\nabla}}\Tilde{\phi},
\end{equation}

\begin{equation}
\label{eq:SF_momentum_conservation_sc}
\begin{split}
    \frac{\partial \Tilde{\bm{j}}}{\partial \Tilde{t}} & + \Tilde{\bm{\nabla}} \Tilde{P} + \Tilde{\rho}_s (\Tilde{\bm{v}}_s\cdot \Tilde{\bm{\nabla}}) \Tilde{\bm{v}}_s + \Tilde{\rho}_n (\Tilde{\bm{v}}_n \cdot \Tilde{\bm{\nabla}}) \Tilde{\bm{v}}_n \\
    & + \Tilde{\bm{v}}_s [\Tilde{\bm{\nabla}} \cdot (\Tilde{\rho}_s \Tilde{\bm{v}}_s)] + \Tilde{\bm{v}}_n [\Tilde{\bm{\nabla}} \cdot( \Tilde{\rho}_n \Tilde{\bm{v}}_n)] =  -\Tilde{\rho}\Tilde{\bm{\nabla}}\Tilde{\phi},
\end{split}
\end{equation}

\begin{equation}
\label{eq:SF_energy_conservation_sc}
\begin{split}
    \frac{\partial \Tilde{E}}{\partial \Tilde{t}} & + \Tilde{\bm{\nabla}}\cdot\Big[(\Tilde{U} + \frac{1}{2}\Tilde{\rho}_n \Tilde{v}_n^2 + \Tilde{P})\Tilde{\bm{v}}_n + \frac{1}{2}\Tilde{\rho}_s \Tilde{v}_s^2 \Tilde{\bm{v}}_s \\
    & + \Tilde{\mu} \Tilde{\rho}_s(\Tilde{\bm{v}}_s - \Tilde{\bm{v}}_n)\Big] = - \Tilde{H}(3\Tilde{P}-2\Tilde{U}) -\Tilde{\bm{j}}\cdot\Tilde{\bm{\nabla}}\Tilde{\phi}.
\end{split}
\end{equation}

The super-comoving quantities are re-scaled to reduce the dependence on the scale factor $a$, with the variables defined as before: $\Tilde{\bm{j}} = \Tilde{\rho}_n \Tilde{\bm{v}}_n + \Tilde{\rho}_s \Tilde{\bm{v}}_s$ and $\Tilde{E} = \Tilde{U} + \frac{1}{2}\Tilde{\rho}_n \Tilde{v}^2_n + \frac{1}{2}\Tilde{\rho}_s \Tilde{v}^2_s$. The only real difference is the peculiar gravitational potential $\Tilde{\phi}$ that is now given by (in a flat universe with matter and a cosmological constant)
\begin{equation}
    \Tilde{\nabla}^2\Tilde{\phi}=6a\left(\Tilde{\rho}-1\right).
\end{equation}
$\Tilde{H}$ is the super-comoving Hubble parameter.

\subsection{Linear perturbation expansion}
\label{sec:lin_pert}
The superfluid hydrodynamic equations at linear order can tell us a lot about the collapse of a superfluid, in particular how it will differ from CDM and non-superfluid thermal DM. The fluid variables are expanded around their background values, $\Tilde{\rho} = \Tilde{\rho}_0 + \delta \Tilde{\rho}$, $\Tilde{S} = \Tilde{S}_0 + \delta \Tilde{S,}$ etc. The peculiar background velocities are zero, so $\Tilde{\bm{v}}_s = \delta \Tilde{\bm{v}}_s$, $\Tilde{\bm{v}}_n = \delta \Tilde{\bm{v}}_n$, and $\Tilde{\bm{j}} = \delta \Tilde{\bm{j}}$. We also have $\Tilde{\nabla}^2 \delta \Tilde{\phi} = 6a\delta\Tilde{\rho}$. This gives the following linear equations;

\begin{equation}
\label{eq:SF_mass_conservation_sc_pert}
    \frac{\partial \delta\Tilde{\rho}}{\partial\Tilde{t}} + \Tilde{\bm{\nabla}} \cdot \delta\Tilde{\bm{j}} = 0,
\end{equation}

\begin{equation}
\label{eq:SF_entropy_conservation_sc_pert}
    \frac{\partial \delta\Tilde{S}}{\partial\Tilde{t}} + \Tilde{S}_0\Tilde{\bm{\nabla}} \cdot \delta\Tilde{\bm{v}}_n = 0,
\end{equation}

\begin{equation}
\label{eq:SF_velocity_sc_pert}
    \frac{\partial \delta\Tilde{\bm{v}}_s}{\partial\Tilde{t}} + \Tilde{\bm{\nabla}} (\delta \Tilde{\mu} + \delta\Tilde{\phi}) = \bm{0},
\end{equation}

\begin{equation}
\label{eq:SF_momentum_conservation_sc_pert}
    \frac{\partial \delta\Tilde{\bm{j}}}{\partial\Tilde{t}} + \Tilde{\bm{\nabla}} \delta \Tilde{P} + \Tilde{\rho_0}\Tilde{\bm{\nabla}}\delta\Tilde{\phi} = \bm{0}.
\end{equation}
These can be combined into two coupled equations for $\delta\Tilde{\rho}$ and $\delta\Tilde{S}$ in $\Tilde{k}$-space;
\begin{equation}
\label{eq:rho_k_mode_pert}
    \frac{\partial^2 \delta\Tilde{\rho}_{\Tilde{k}}}{\partial \Tilde{t}^2} + \Bigg[ \left(\frac{\partial \Tilde{P}}{\partial \Tilde{\rho}}\right)_0 \Tilde{k}^2 - 6a\Tilde{\rho}_0 \Bigg]\delta\Tilde{\rho}_{\Tilde{k}} + \left(\frac{\partial \Tilde{P}}{\partial \Tilde{S}}\right)_0 \Tilde{k}^2 \delta\Tilde{S}_{\Tilde{k}} = 0,
\end{equation}

\begin{equation}
\label{eq:S_k_mode_pert}
\begin{split}
    \frac{\partial^2 \delta\Tilde{S}_{\Tilde{k}}}{\partial \Tilde{t}^2} & + \Tilde{S}_0\Bigg[\Bigg\{ \frac{1}{\Tilde{\rho}_0}\left(\frac{\partial \Tilde{P}}{\partial \Tilde{\rho}}\right)_0 + \frac{\Tilde{S}_0\Tilde{\rho}_{s0}}{\Tilde{\rho}_{0}\Tilde{\rho}_{n0}} \left(\frac{\partial \Tilde{T}}{\partial \Tilde{\rho}}\right)_0 \Bigg\} \Tilde{k}^2 - 6a \Bigg] \delta\Tilde{\rho}_{\Tilde{k}} \\
     & + \Tilde{S}_0\Bigg[ \frac{1}{\Tilde{\rho}_0}\left(\frac{\partial \Tilde{P}}{\partial \Tilde{S}}\right)_0 + \frac{\Tilde{S}_0\Tilde{\rho}_{s0}}{\Tilde{\rho}_{0}\Tilde{\rho}_{n0}} \left(\frac{\partial \Tilde{T}}{\partial \Tilde{S}}\right)_0 \Bigg] \Tilde{k}^2 \delta\Tilde{S}_{\Tilde{k}} = 0,
\end{split}
\end{equation}
where the subscript "$0$" indicates the background values.

We would like to enforce the critical velocity in the linear approach, though we cannot do it in the same way as for the full hydrodynamic equations. Since the effect of the critical velocity is to essentially restrict the two-fluid nature of the superfluid, forcing the whole fluid to evolve like a normal fluid, we can, as a rough approximation, set $\rho_s=0$ and $\rho_n = \rho$ once $\Tilde{w}_{\Tilde{k}} \geq \Tilde{v}_c$, where $\Tilde{w}_{\Tilde{k}}$ is the relative velocity of mode $\Tilde{k}$ and evolves at linear order according to
\begin{equation}
    \frac{\partial \Tilde{w}_{\Tilde{k}}}{\partial \Tilde{t}} = \Tilde{k}\frac{\Tilde{S}_0}{\Tilde{\rho}_{n0}}\Bigg[\left(\frac{\partial \Tilde{T}}{\partial \Tilde{\rho}}\right)_0\delta\Tilde{\rho}_{\Tilde{k}} + \left(\frac{\partial \Tilde{T}}{\partial \Tilde{S}}\right)_0\delta\Tilde{S}_{\Tilde{k}} \Bigg].
\end{equation}
This approximation is further justified by the fact that the critical velocity decreases with the DM density. Once $w$ reaches $v_c,$ it only becomes smaller in the linear regime, forcing the superfluid to behave even more like a normal fluid.

A few qualitative statements can be made from Eqs. \eqref{eq:rho_k_mode_pert} and \eqref{eq:S_k_mode_pert}. Both mass density and entropy perturbations grow due to gravity, but this growth is slowed by pressure terms that are scale dependent through the $\Tilde{k}^2$ factor, as expected in a self-gravitating fluid with nonzero pressure. In a superfluid, however, there are additional effective pressure terms that suppress the growth of entropy perturbations, and hence thermal pressure, that are absent in conventional fluids. This in turn allows the mass density perturbations to collapse more efficiently, even though the DM fluid may have relatively high temperatures. The reason for this behavior is the superfluid component's attraction to higher temperatures. The normal component tends to transport mass and entropy from hot to cold regions, while the superfluid tends to flow in the opposite direction and balance the mass-loss due to the normal component, resulting in a thermal flux that can be large compared to the net mass flux. This effect, called thermal counterflow, makes superfluids very efficient at conducting heat.

\section{Results and discussion}
\label{sec:results}

\begin{figure}%[H]
    \centering
    \includegraphics[width=1\linewidth]{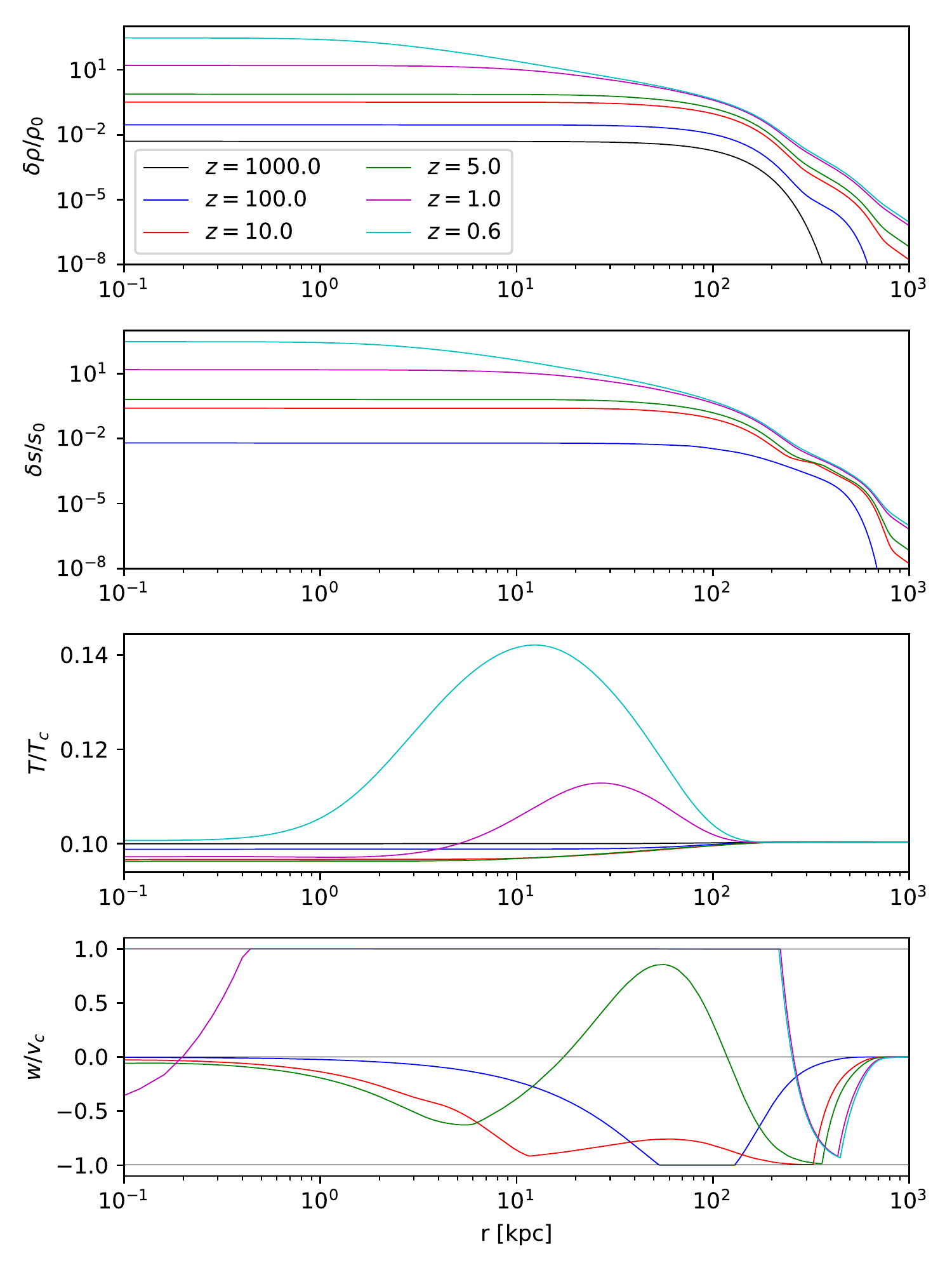}
    \caption{Profiles of a collapsing SFDM halo with an initial Gaussian density contrast, $m=30\,\text{eV}$, $g=10^{-5}\,\text{eV}^{-2}$, $L=100\,\text{kpc}$, and $T/T_c=0.1$. A thermal counterflow develops and the growth of entropy perturbations is at first suppressed. This also gives a slight decrease in the ratio $T/T_c$, and hence the superfluid fraction, since $f_s=\rho_s/\rho = 1-(T/T_c)^{3/2}$. As the critical velocity is reached, entropy is generated, and $T/T_c$ increases.}
    \label{fig:halo_profiles_at_redshifts}
\end{figure}

The hydrodynamic equations were integrated numerically using a modified first-order FORCE scheme (see \citet{Toro2006} and Appendix \ref{app:numerical_scheme} for further details) for a spherically symmetric system with an initial density contrast of the form $\delta\Tilde{\rho}/\Tilde{\rho}_0 = \Delta_0 e^{-(\Tilde{r}/\Tilde{L})^2}$ and $\delta\Tilde{S} = 0$, where $\Tilde{L}$ is the size of the overdensity. The initial state is at approximately the same $T/T_c$, and hence the same mixing fraction of the normal and superfluid components, throughout the system. A flat $\Lambda$CDM background cosmology with $\Omega_{m0} = 0.3$, $\Omega_{\Lambda0} = 0.7$, and $h=0.7$ was used, and the integration started at redshift $z=1000$ with $\Delta_0=5\times10^{-3}$. An example of a collapsing SFDM halo at various redshifts can be seen in \figref{fig:halo_profiles_at_redshifts}, illustrating that as the halo collapses, a thermal counterflow carrying entropy away from the halo center develops, slowing down the growth of entropy until the critical velocity is reached and the fluid starts heating up.

\subsection{Growth of structure}

\begin{figure}%[H]
    \centering
    \includegraphics[width=1\linewidth]{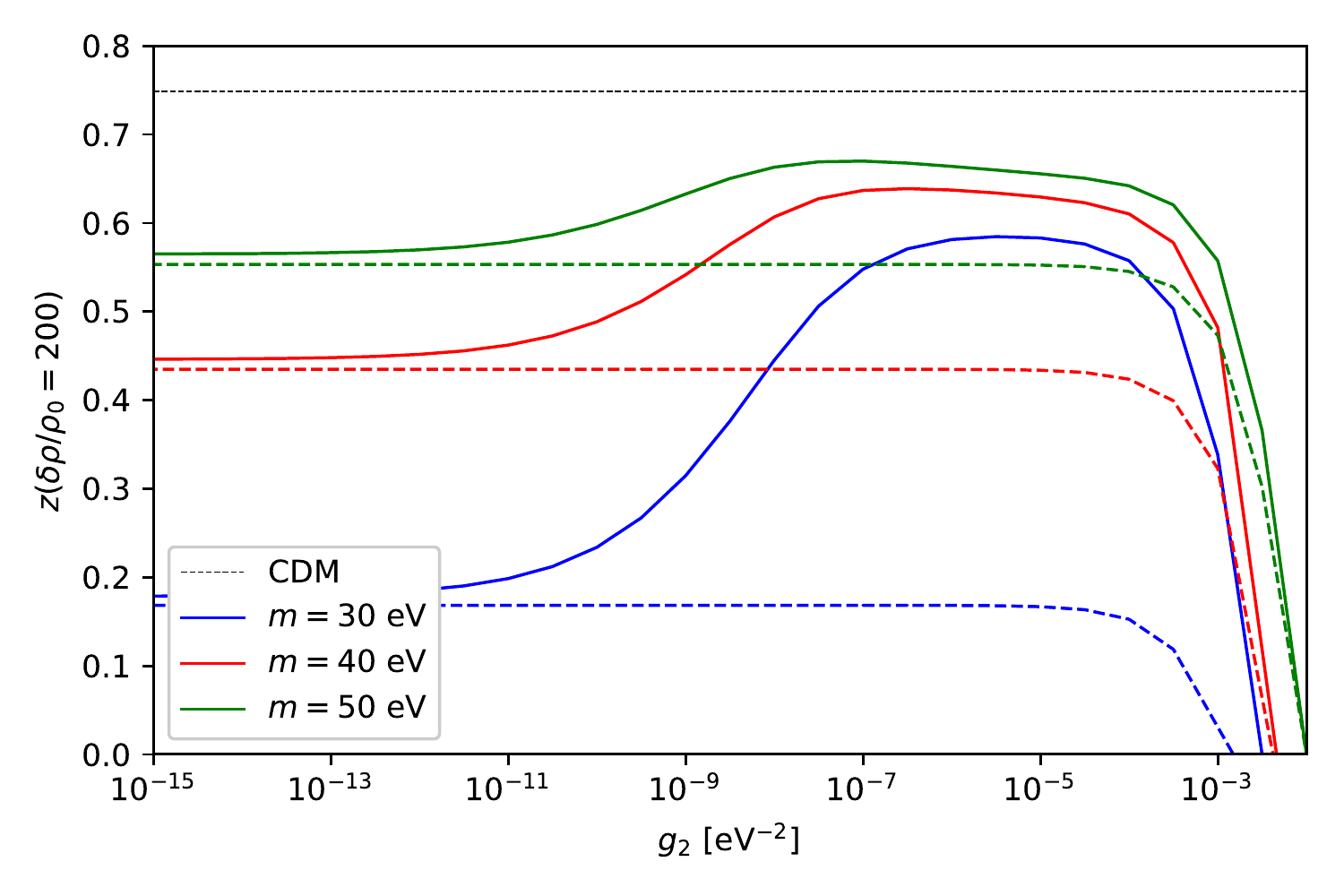}
    \caption{Comparison of redshifts when the central density contrast reaches 200 as function of the interaction strength for various particle masses, with $T/T_c = 0.1$ and $L=100\,\text{kpc}$. Both the superfluid case (solid lines) and the corresponding non-superfluid case (striped lines) are shown. For constant $T/T_c,$ the temperature is increased for decreasing mass, since $T_c \sim m^{-5/3}$. The comparison of the collapse for various masses is therefore not done at the same temperature, but instead at a similar place in the superfluid phase.}
    \label{fig:mass_collapse_comparison}
\end{figure}

\begin{figure}%[H]
    \centering
    \includegraphics[width=1\linewidth]{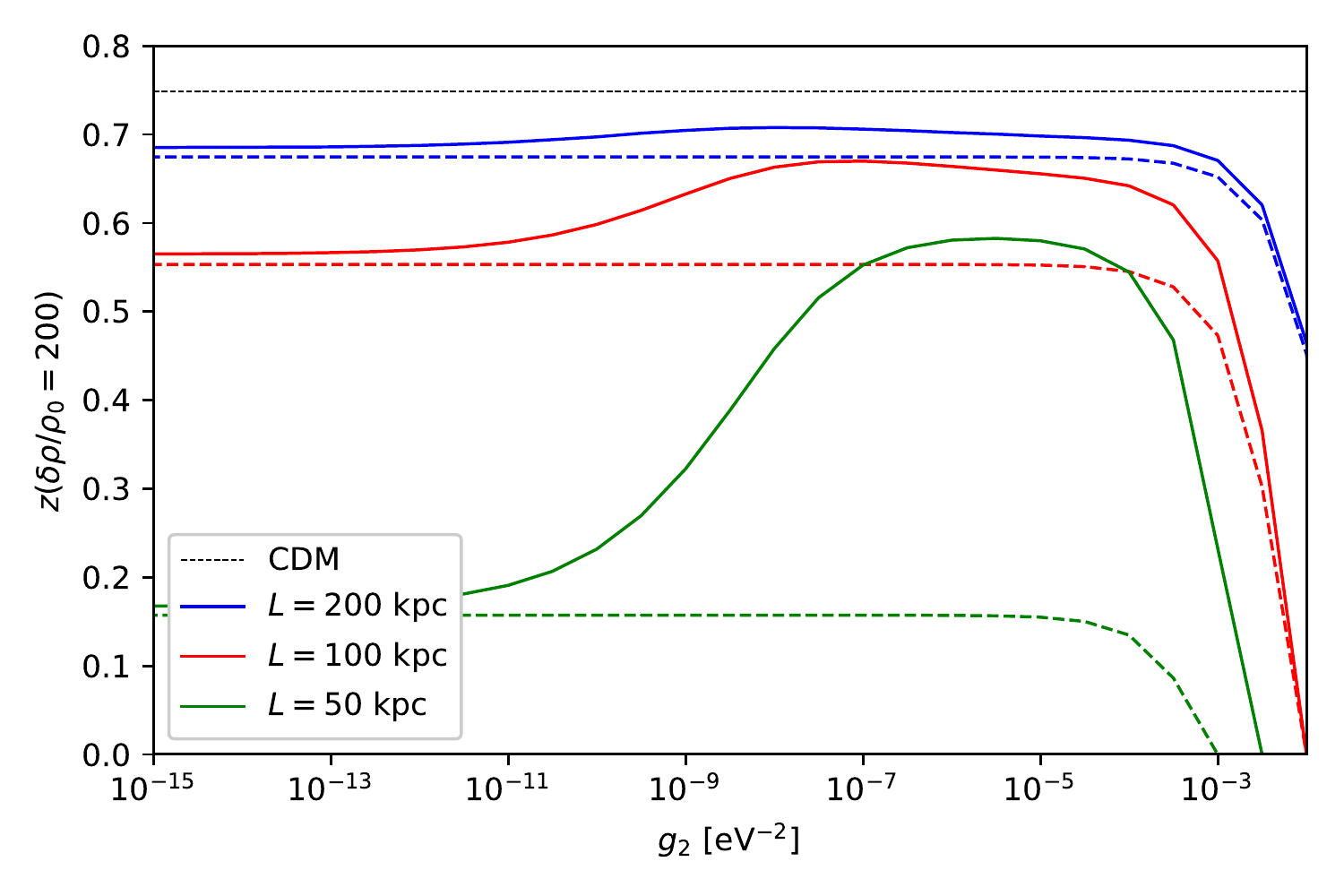}
    \caption{Comparison of redshifts when the central density contrast reaches 200 as function of the interaction strength for various scales, with $T/T_c = 0.1$ and $m=50\,\text{eV}$. Both the superfluid case (solid lines) and the corresponding non-superfluid case (striped lines) are shown.}
    \label{fig:scale_collapse_comparison}
\end{figure}

\begin{figure}%[H]
    \centering
    \includegraphics[width=1\linewidth]{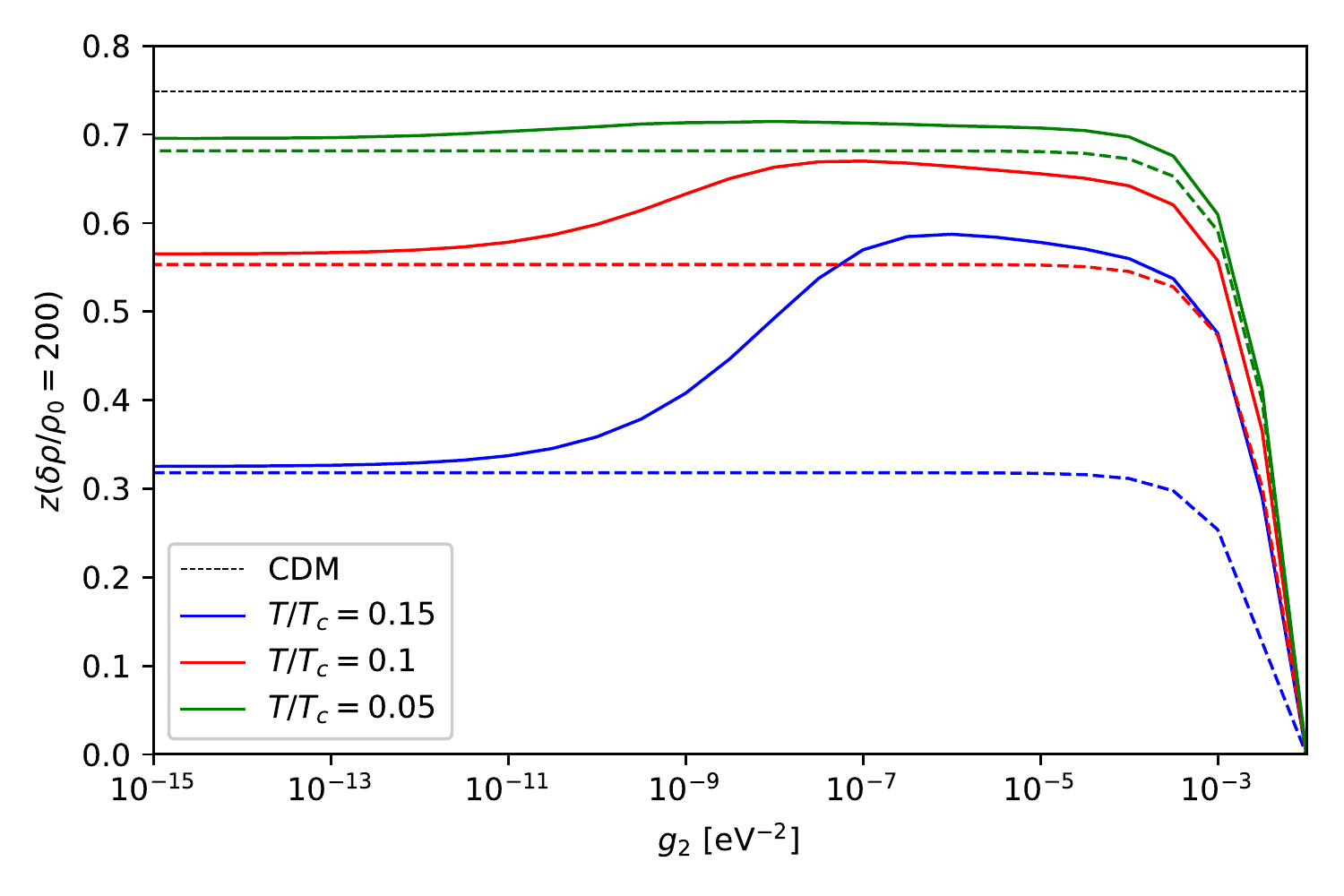}
    \caption{Comparison of redshifts when the central density contrast reaches 200 as function of the interaction strength for various temperatures, with $m=50\,\text{eV}$ and $L=100\,\text{kpc}$. Both the superfluid case (solid lines) and the corresponding non-superfluid case (striped lines) are shown.}
    \label{fig:temperature_collapse_comparison}
\end{figure}

In \figref{fig:mass_collapse_comparison}, \figref{fig:scale_collapse_comparison}, and \figref{fig:temperature_collapse_comparison}, the redshift when the central density contrast reaches $200$ is shown for various parameters for both the superfluid and non-superfluid (a conventional fluid with $\rho_s=0$, $\rho_n = \rho$, and the same EOS) cases. While the growth of structure is slower compared to CDM, the SFDM halos collapse more efficiently than their non-superfluid counterparts as the interaction strength is increased until a maximum is reached, after which the growth of structure in both super-and non-superfluid DM is suppressed. This is counter to what one would intuitively expect, since an increase in interactions also means an increase in pressure. It can, however, be understood as follows: for small interactions, the superfluid behaves nearly the same as a normal fluid because the critical velocity, which scales as $v_c \sim \sqrt{g_i}$, is reached very early. When this happens, the flow of the normal and superfluid components become "locked" to one another, unable to efficiently conduct heat away from the halo core. As the interaction increases, the thermal counterflow can both be larger and last longer, resulting in an increased suppression of thermal gradients and thus allows for a faster collapse. For sufficiently large interactions, the collapse is instead suppressed due to large zero-temperature pressure gradients that the superfluid is unable to wash out.

Most production of entropy due to mutual friction as the Landau criterion is broken takes place away from the center of the halo. The resulting extra thermal pressure acting on the interior causes the central density contrast to grow slightly faster and can best be seen by the gap between collapse times of the superfluid and non-superfluid cases at low $g_2$. If entropy was not produced, this gap would vanish.

\subsection{Dependence on equation of state}

\begin{figure}%[H]
    \centering
    \includegraphics[width=1\linewidth]{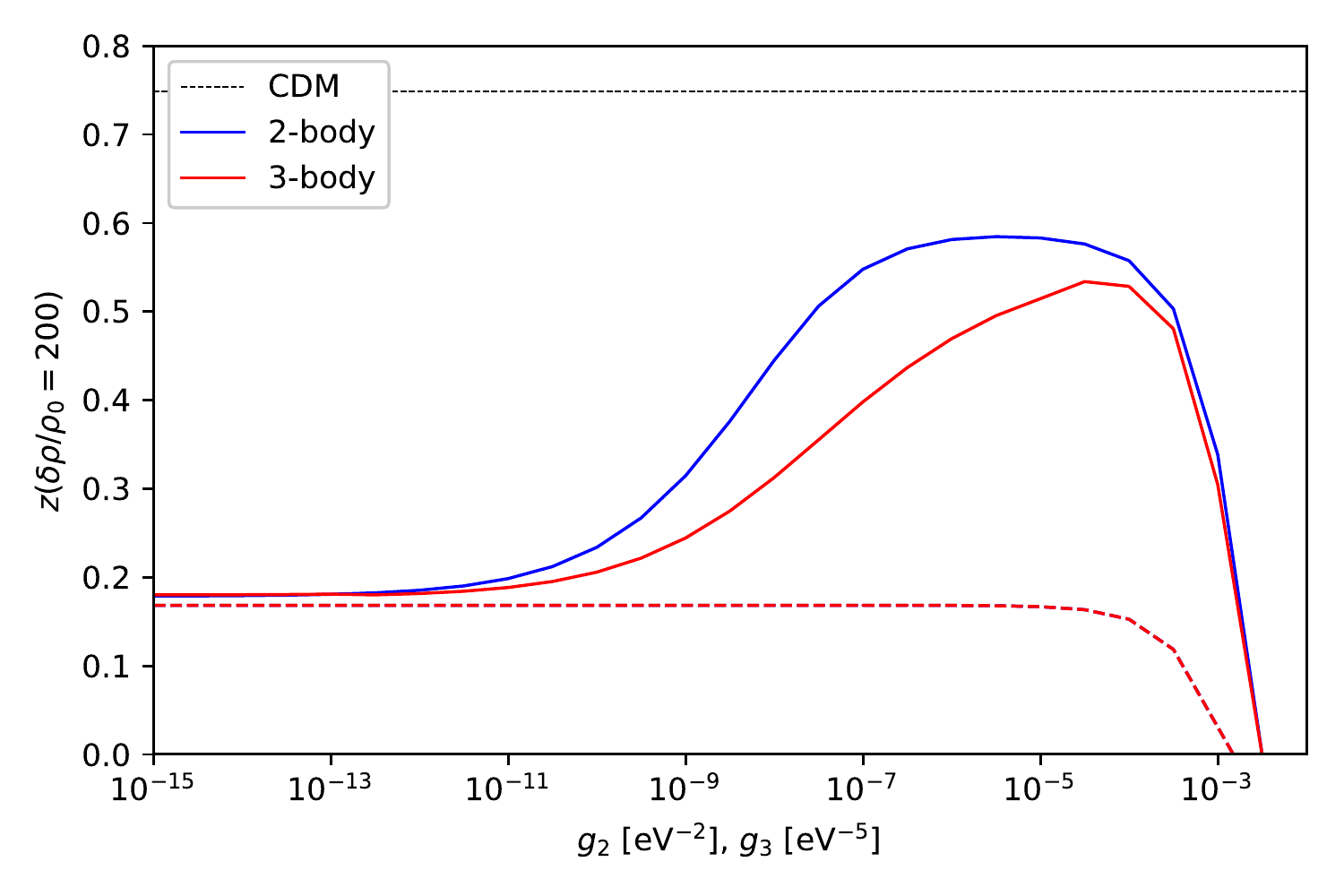}
    \caption{Comparison of redshifts when the central density contrast reaches 200 for two-body and three-body interactions as function of the interaction strengths $g_2$ and $g_3,$ respectively, with $m=30\,\text{eV}$, $L=100\,\text{kpc}$, and $T/T_c = 0.1$. The three-body interaction is multiplied by $\sqrt{10}\times 10^{-4}$ to make the comparison clearer. Both the superfluid case (solid lines) and the non-superfluid (striped lines) are shown.}
    \label{fig:2b_3b_collapse_comparison}
\end{figure}

The Bose gas with two-body interactions is compared with three-body interactions in \figref{fig:2b_3b_collapse_comparison}. The same qualitative behavior is present in both cases and is expected to be a general feature regardless of the EOS used, as long as there is superfluidity. In the linear expansion of the superfluid equations, Eqs. $\eqref{eq:rho_k_mode_pert}$ and $\eqref{eq:S_k_mode_pert}$, the additional effective pressure terms due to a superfluid component require only the temperature to be dependent on mass density or entropy. Indeed, the approximated two-body and three-body EOS used in this work both have a temperature profile that is independent of mass density for $T<T_c$, so that one of the effective pressure terms in Eq. \eqref{eq:S_k_mode_pert} is absent. For EOS where the temperature is dependent on both the mass density and entropy, the collapse of SFDM may be even more efficient.

\subsection{Effect of small-scale and nonradial motion}

\begin{figure}%[H]
    \centering
    \includegraphics[width=1\linewidth]{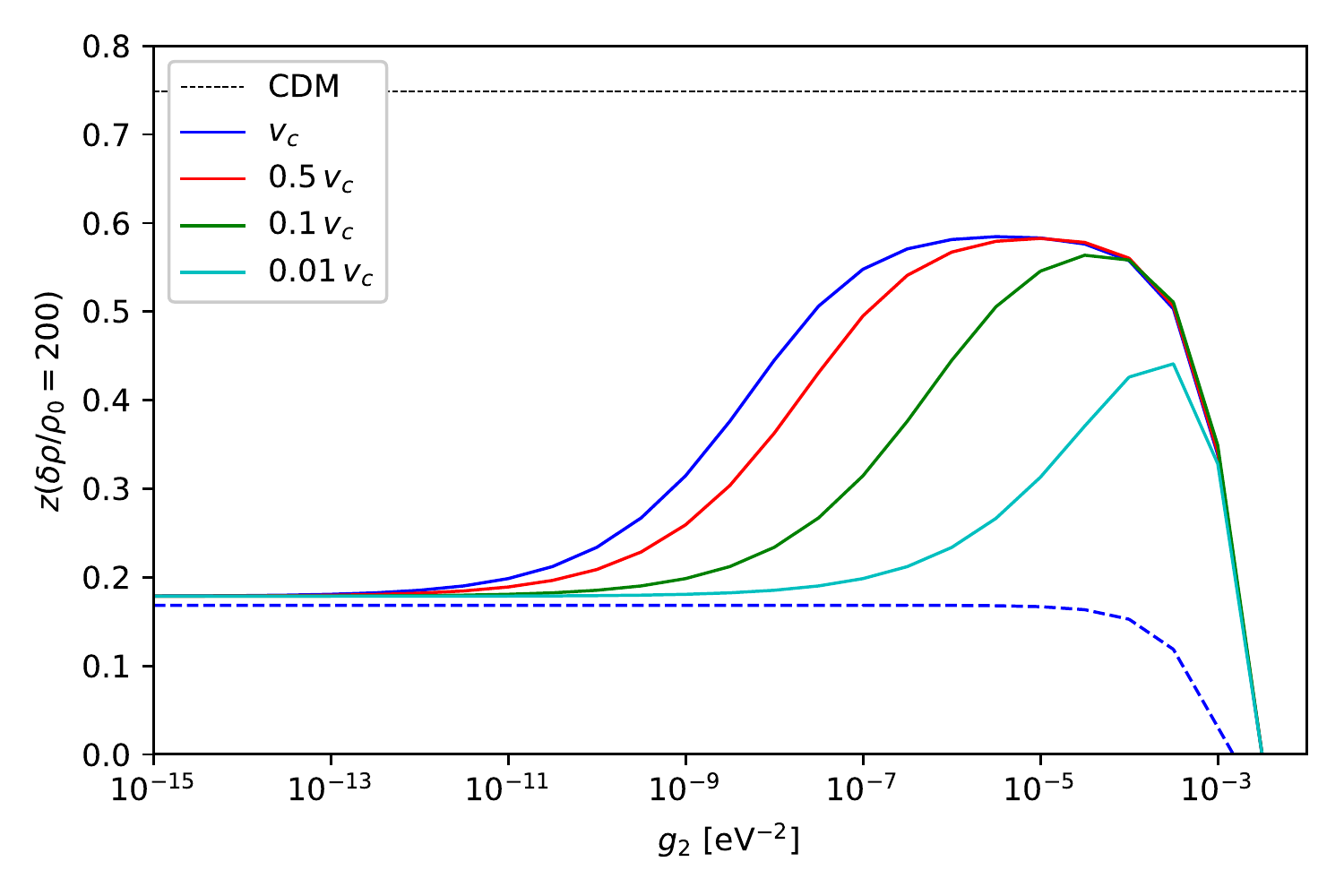}
    \caption{Comparison of redshifts when the central density contrast reaches 200 for various effective critical velocities as function of the interaction strength, with $m=30\,\text{eV}$, $T/T_c=0.1$, and $L=100\,\text{kpc}$. Both the superfluid case (solid lines) and the non-superfluid (striped line) are shown.}
    \label{fig:vc_eff_collapse_comparison}
\end{figure}

In this work, we assumed perfect radial infall of DM. The relative velocity $w$ is simply the difference between the radial velocities of the two fluid components. In a real system, there is expected to be additional small-scale motion in all directions, such as turbulence that our simplified model averages over. The superfluid critical velocity may therefore be exceeded on small scales, while the large-scale radial average only appears to have $w<v_c$. In this case, the superfluid would behave like a conventional fluid at much smaller $w$. In other words, there is an effective superfluid critical velocity $v_c^{\text{eff}} < v_c$ that is a decreasing function of the local turbulence. This leads to a difference in collapse times of halos with different amounts of turbulence, the turbulent ones collapsing at a slower rate, as seen in \figref{fig:vc_eff_collapse_comparison}.

\subsection{Evolution of superfluid fraction}

\begin{figure}%[H]
    \centering
    \includegraphics[width=1\linewidth]{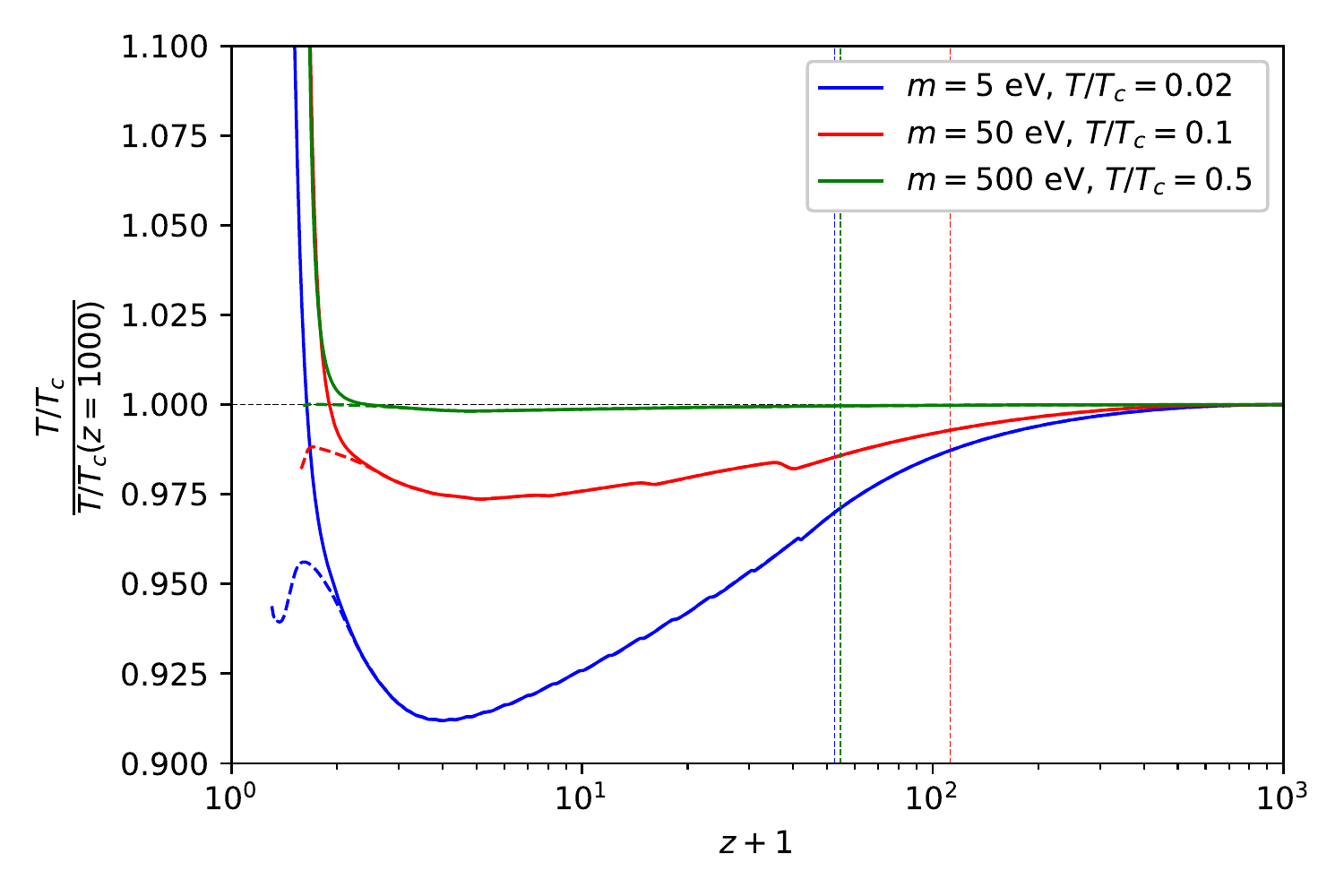}
    \caption{Evolution of $T/T_c$ in the halo center during collapse for various masses and initial temperatures with $g=10^{-5}\,\text{eV}^{-2}$ and $L=100\,\text{kpc}$. Both the evolution with entropy production (solid lines) and without (striped lines) are shown until the overdensity reach $10^{5}$. The two cases differ only in the end stage of the collapse, well after the critical velocity is first reached, indicated by the colored vertical lines.}
    \label{fig:TdTc_evolution}
\end{figure}

\begin{figure}%[H]
    \centering
    \includegraphics[width=1\linewidth]{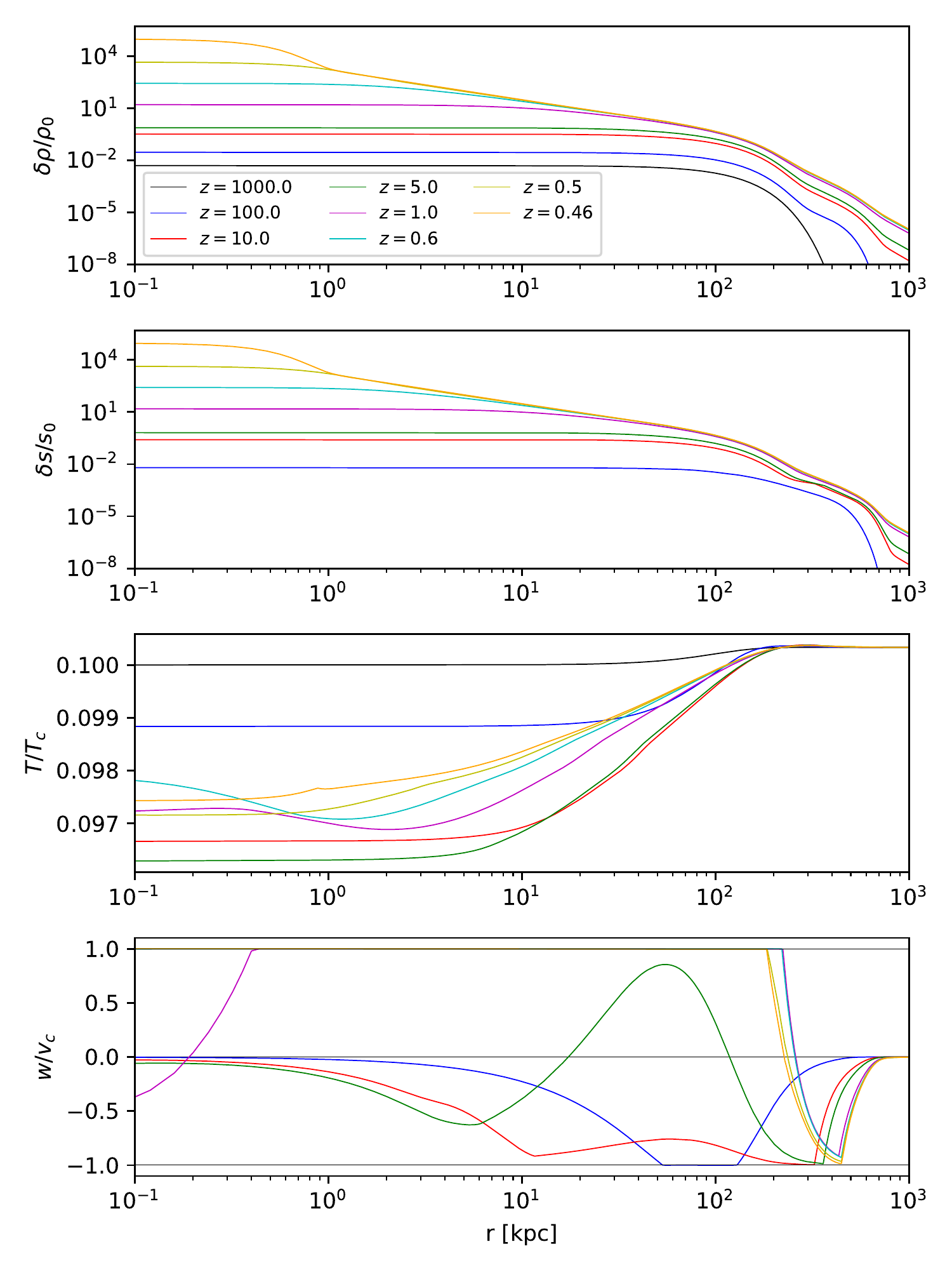}
    \caption{Profiles of a collapsing SFDM halo where no entropy is produced as $w=v_v$ with an initial gaussian density contrast, $m=30\,\text{eV}$, $g=10^{-5}\,\text{eV}^{-2}$, $L=100\,\text{kpc}$, and $T/T_c=0.1$.}
    \label{fig:halo_profiles_at_redshifts_no_entr_prod}
\end{figure}

In a conventional fluid, the entropy and mass density collapses at the same rate so that the ratio $T/T_c$ is constant. A fluid that is initially in the normal phase will therefore remain so. A collapsing superfluid, on the other hand, experiences an increase in the superfluid fraction due to thermal counterflow until the critical velocity is reached. At this point, entropy is generated causing $T/T_c$ to rise, and thus the superfluid fraction to fall; though it takes time for the full effect of this to propagate to the center of the halo, as shown in \figref{fig:halo_profiles_at_redshifts} and \figref{fig:TdTc_evolution}. It may be, however, that Eqs. \eqref{eq:SF_mass_conservation}-\eqref{eq:SF_energy_conservation} do not properly describe super-critical flow, and too much entropy is generated in our numerical scheme for enforcing the critical velocity. The evolution of $T/T_c$ when no entropy is generated is therefore also shown in \figref{fig:TdTc_evolution} as the opposite extreme. This case behaves similarly until near the end of the collapse, where $T/T_c$ rises only modestly. Profiles are shown in \figref{fig:halo_profiles_at_redshifts_no_entr_prod}, which corresponds to \figref{fig:halo_profiles_at_redshifts} with no production of entropy.

The decrease in $T/T_c$ during collapse becomes smaller as the temperature approaches $T_c,$ where the superfluid fraction goes to zero and thermal counterflow becomes inefficient. The formation of DM halos with much higher superfluid fractions than the background, as required in the emergent MOND scenario of \citet{Berezhiani2015}, therefore appears unlikely through collapse alone. Additional cooling mechanisms during or after collapse are necessary. 

\subsection{Dark matter self-interaction constraints}

\begin{figure}%[H]
    \centering
    \includegraphics[width=1\linewidth]{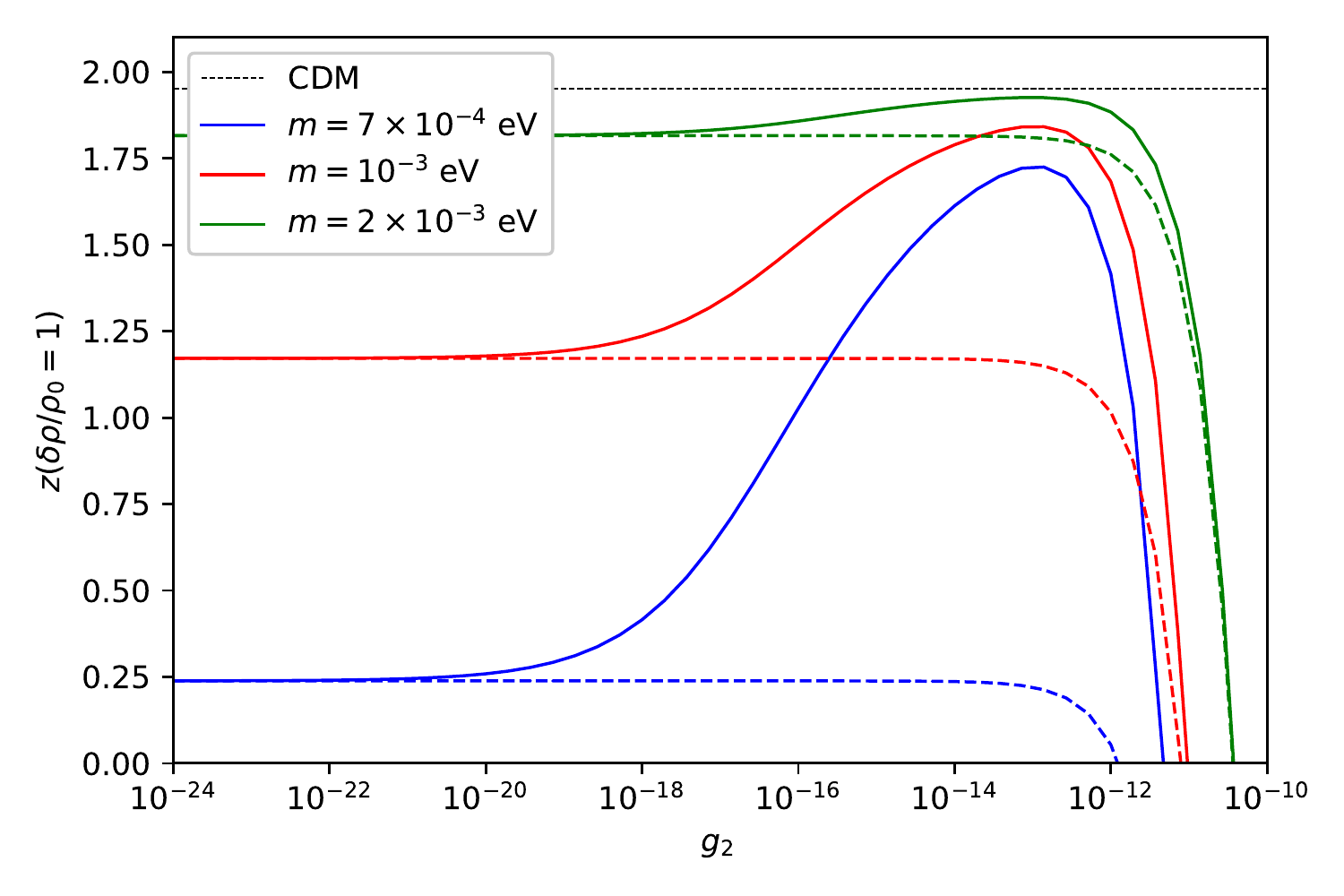}
    \caption{Redshifts when the linear density contrast for the mode $k=2/100\,\text{kpc}^{-1}$ with $T/T_c = 2\times 10^{-6}$ reaches unity for various masses and interaction strengths. Both the superfluid case (solid lines) and the corresponding non-superfluid case (striped lines) are shown, illustrating that the same features can be found for a choice of parameters that satisfy the constraint from cluster collisions on DM mass and self-interaction.}
    \label{fig:mass_collapse_comparison_linear}
\end{figure}

The distribution of DM, gas, and stellar mass in cluster collisions provides constraints on the cross-section of DM self-interactions, $\sigma/m < 0.5\,\text{cm}^2\text{/g}$ \citep{Harvey2015}. In terms of the two-body interaction strength, this corresponds to \citep{Pitaevskii2016}
\begin{equation}
    g_2 = \sqrt{4\pi\sigma}\frac{\hbar^2}{m} < 5\times 10^{-12}\left(\frac{1\,\text{eV}}{m}\right)^{1/2} \,\text{eV}^{-2}.
\end{equation}
The values of $g_2$ in the above results do not generally satisfy this constraint, but we chose to relax it since we do not know how it translates to SFDM. In any case, the above features were also found for smaller $g_2$ using perturbation theory (while simultaneously lowering $m$ and $T/T_c$) that do satisfy the constraints, as is exemplified in \figref{fig:mass_collapse_comparison_linear}.

\section{Conclusions}
\label{sec:conclusions}

When superfluid behavior is included in a finite-temperature DM fluid, the formation of structure is found to be much more efficient in certain regions of parameter space than one would, naively, expect, through it is still slower compared to CDM. The effect of thermal counterflow is most prominent when the thermal suppression is large, such as at small scales and relatively high temperatures. The increased collapse efficiency is also expected to be a general feature of SFDM regardless of the EOS used, though the specific model in question will certainly affect the finer details through the dependence of entropy, pressure, and critical velocity on temperature, mass density, and the model parameters. The toy models used in this work were motivated by condensed matter physics, but suffer some severe limitations at high redshifts. Both are derived under the assumption that the interactions are weak and the number density is not too large, which is invalid at very early times. Furthermore, the zero-temperature pressure depends on the number density through $n^2$ and $n^3$, resulting in very high pressures at high redshifts that might wash out the initial perturbations set up by inflation. The generalization of this work to more exotic DM fluids and adding interactions between DM and baryons, which has recently been considered in the literature, is therefore of interest in the further study of SFDM models. It may also be of interest to study the case when thermal equilibrium is not always assumed so that the DM fluid can fall in and out of equilibrium and superfluidity can vanish and reappear.

Superfluid models of DM involve processes that require the superfluid hydrodynamic equations to be properly described. Throughout this work, spherical symmetry was assumed, but nonradial and turbulent motion is expected to have a significant impact on the superfluid dynamics, especially through the critical velocity, which is broken at smaller radial thermal counterflows. It is also important to understand the effect of mergers in SFDM. Large-scale and high-resolution simulations will therefore be essential for the further study of structure formation. The main challenge in this line of inquiry may be developing numerical schemes that are faster and more accurate than the modified first-order FORCE scheme used in this work that can capture the small-scale motion of the superfluid and its effect on structure formation.

\begin{acknowledgements}
We thank the Research Council of Norway for their support. We are grateful for the helpful comments and suggestions made by the anonymous referee.
\end{acknowledgements}

%%%%%%%%%%%%%%%%%%%%%%%%%%%%%%%%%%%%%%%%%%%%%%%%%%%%%%%%%%%%%%%%%%%%%%%%%%%%%%%%%
%%%%%%%%%%%%%%%%%%%%%%%%%%%%%%%%%%%%%%%%%%%%%%%%%%%%%%%%%%%%%%%%%%%%%%%%%%%%%%%%%
%%%%%%%%%%%%%%%%%%%%%%%%%%%%%%%%%%%%%%%%%%%%%%%%%%%%%%%%%%%%%%%%%%%%%%%%%%%%%%%%%
%%%%%%%%%%%%%%%%%%%%%%%%%%%%%%%%%%%%%%%%%%%%%%%%%%%%%%%%%%%%%%%%%%%%%%%%%%%%%%%%%

%%%%%%%%%%%%%%%%%%%%%%%%%%%%%%%%%%%%%%%%%%%
%%%%%%%%%%%%%%%%%%%%%%%%%%%%%%%%%%%%%%%%%%%
%%%%%% Generate references from refs.bib
%\bibliographystyle{aa} % style aa.bst
%\bibliography{refs} % use refs.bib for the references

\begin{thebibliography}{49}
\expandafter\ifx\csname natexlab\endcsname\relax\def\natexlab#1{#1}\fi

\bibitem[{Andersen(2004)}]{Andersen2004}
Andersen, J.~O. 2004, Rev. Mod. Phys., 76, 599

\bibitem[{Angus {et~al.}(2014)Angus, Diaferio, Famaey, Gentile, \& van~der
  Heyden}]{Angus2014}
Angus, G., Diaferio, A., Famaey, B., Gentile, G., \& van~der Heyden, K. 2014,
  Journal of Cosmology and Astroparticle Physics, 2014, 079

\bibitem[{Angus {et~al.}(2013)Angus, Diaferio, Famaey, \& van~der
  Heyden}]{Angus2013}
Angus, G.~W., Diaferio, A., Famaey, B., \& van~der Heyden, K.~J. 2013, Monthly
  Notices of the Royal Astronomical Society, 436, 202

\bibitem[{Barenghi {et~al.}(2014)Barenghi, Skrbek, \&
  Sreenivasan}]{Barenghi2014}
Barenghi, C.~F., Skrbek, L., \& Sreenivasan, K.~R. 2014, Proceedings of the
  National Academy of Sciences, 111, 4647

\bibitem[{Berezhiani \& Khoury(2015)}]{Berezhiani2015}
Berezhiani, L. \& Khoury, J. 2015, Phys. Rev. D, 92, 103510

\bibitem[{Bullock \& Boylan-Kolchin(2017)}]{Bullock2017}
Bullock, J.~S. \& Boylan-Kolchin, M. 2017, Annual Review of Astronomy and
  Astrophysics, 55, 343

\bibitem[{Chapman {et~al.}(2014)Chapman, Hoyos, \& Oz}]{Chapman2014}
Chapman, S., Hoyos, C., \& Oz, Y. 2014, Journal of High Energy Physics, 2014,
  27

\bibitem[{Cyburt {et~al.}(2016)Cyburt, Fields, Olive, \& Yeh}]{Cyburt2016}
Cyburt, R.~H., Fields, B.~D., Olive, K.~A., \& Yeh, T.-H. 2016, Rev. Mod.
  Phys., 88, 015004

\bibitem[{Darve {et~al.}(2012)Darve, Bottura, Patankar, \&
  Van~Sciver}]{Darve2012}
Darve, C., Bottura, L., Patankar, N.~A., \& Van~Sciver, S. 2012, AIP Conference
  Proceedings, 1434, 247

\bibitem[{Del~Popolo \& Le~Delliou(2017)}]{DelPopolo2017}
Del~Popolo, A. \& Le~Delliou, M. 2017, Galaxies, 5

\bibitem[{Dodelson(2011)}]{Dodelson2011}
Dodelson, S. 2011, International Journal of Modern Physics D, 20, 2749

\bibitem[{Doi {et~al.}(2008)Doi, Shirai, \& Shiotsu}]{Doi2008}
Doi, D., Shirai, Y., \& Shiotsu, M. 2008, AIP Conference Proceedings, 985, 648

\bibitem[{Elbert {et~al.}(2015)Elbert, Bullock, Garrison-Kimmel, Rocha,
  Oñorbe, \& Peter}]{Elbert2015}
Elbert, O.~D., Bullock, J.~S., Garrison-Kimmel, S., {et~al.} 2015, Monthly
  Notices of the Royal Astronomical Society, 453, 29

\bibitem[{Famaey \& McGaugh(2012)}]{Famaey2012}
Famaey, B. \& McGaugh, S.~S. 2012, Living Reviews in Relativity, 15, 10

\bibitem[{Glyde(2013)}]{Glyde2013}
Glyde, H.~R. 2013, Journal of Low Temperature Physics, 172, 364

\bibitem[{Harvey {et~al.}(2015)Harvey, Massey, Kitching, Taylor, \&
  Tittley}]{Harvey2015}
Harvey, D., Massey, R., Kitching, T., Taylor, A., \& Tittley, E. 2015, Science,
  347, 1462

\bibitem[{Hu {et~al.}(2000)Hu, Barkana, \& Gruzinov}]{Hu2000}
Hu, W., Barkana, R., \& Gruzinov, A. 2000, Phys. Rev. Lett., 85, 1158

\bibitem[{Khalatnikov(2000)}]{Khalatnikov2000}
Khalatnikov, I.~M. 2000, An Introduction To The Theory Of Superfluidity, 1st
  edn. (Westview Press)

\bibitem[{Khoury(2016)}]{Khoury2016}
Khoury, J. 2016, Phys. Rev. D, 93, 103533

\bibitem[{Landau(1941)}]{Landau1941}
Landau, L. 1941, Phys. Rev., 60, 356

\bibitem[{Landau \& Lifshitz(1987)}]{Landau1987}
Landau, L.~D. \& Lifshitz, E.~M. 1987, Course of theoretical physics / by L. D.
  Landau and E. M. Lifshitz, Vol. 6, 2nd edn. (Butterworth-Heinemann)

\bibitem[{Lelli {et~al.}(2015)Lelli, McGaugh, \& Schombert}]{Lelli2015}
Lelli, F., McGaugh, S.~S., \& Schombert, J.~M. 2015, The Astrophysical Journal,
  816, L14

\bibitem[{Martel \& Shapiro(1998)}]{Shapiro1998}
Martel, H. \& Shapiro, P.~R. 1998, Mon. Not. R. Astron. Soc, 297, 467

\bibitem[{McGaugh(2005)}]{McGaugh2005}
McGaugh, S.~S. 2005, The Astrophysical Journal, 632, 859

\bibitem[{McGaugh(2012)}]{McGaugh2012}
McGaugh, S.~S. 2012, The Astronomical Journal, 143, 40

\bibitem[{McGaugh {et~al.}(2000)McGaugh, Schombert, Bothun, \&
  de~Blok}]{McGaugh2000}
McGaugh, S.~S., Schombert, J.~M., Bothun, G.~D., \& de~Blok, W. J.~G. 2000, The
  Astrophysical Journal, 533, L99

\bibitem[{Milgrom(1983{\natexlab{a}})}]{Milgrom1983b}
Milgrom, M. 1983{\natexlab{a}}, \apj, 270, 371

\bibitem[{Milgrom(1983{\natexlab{b}})}]{Milgrom1983c}
Milgrom, M. 1983{\natexlab{b}}, \apj, 270, 384

\bibitem[{Milgrom(1983{\natexlab{c}})}]{Milgrom1983a}
Milgrom, M. 1983{\natexlab{c}}, \apj, 270, 365

\bibitem[{Mocz {et~al.}(2017)Mocz, Vogelsberger, Robles, Zavala,
  Boylan-Kolchin, Fialkov, \& Hernquist}]{Mocz2017}
Mocz, P., Vogelsberger, M., Robles, V.~H., {et~al.} 2017, Monthly Notices of
  the Royal Astronomical Society, 471, 4559

\bibitem[{Pethick \& Smith(2008)}]{Pethick2008}
Pethick, C.~J. \& Smith, H. 2008, Bose–Einstein Condensation in Dilute Gases,
  2nd edn. (Cambridge University Press)

\bibitem[{Pitaevskii \& Stringari(2016)}]{Pitaevskii2016}
Pitaevskii, L.~P. \& Stringari, S. 2016, Bose-Einstein Condensation and
  Superfluidity (Great Clarendon Street, Oxford, United Kingdom: Oxford
  University Press)

\bibitem[{{Planck Collaboration} {et~al.}(2016){Planck Collaboration}, {Ade},
  {Aghanim}, {Arnaud}, {Ashdown}, {Aumont}, {Baccigalupi}, {Banday},
  {Barreiro}, {Bartlett}, {Bartolo}, {Battaner}, {Battye}, {Benabed},
  {Beno{\^\i}t}, {Benoit-L{\'e}vy}, {Bernard}, {Bersanelli}, {Bielewicz},
  {Bock}, {Bonaldi}, {Bonavera}, {Bond}, {Borrill}, {Bouchet}, {Boulanger},
  {Bucher}, {Burigana}, {Butler}, {Calabrese}, {Cardoso}, {Catalano},
  {Challinor}, {Chamballu}, {Chary}, {Chiang}, {Chluba}, {Christensen},
  {Church}, {Clements}, {Colombi}, {Colombo}, {Combet}, {Coulais}, {Crill},
  {Curto}, {Cuttaia}, {Danese}, {Davies}, {Davis}, {de Bernardis}, {de Rosa},
  {de Zotti}, {Delabrouille}, {D{\'e}sert}, {Di Valentino}, {Dickinson},
  {Diego}, {Dolag}, {Dole}, {Donzelli}, {Dor{\'e}}, {Douspis}, {Ducout},
  {Dunkley}, {Dupac}, {Efstathiou}, {Elsner}, {En{\ss}lin}, {Eriksen},
  {Farhang}, {Fergusson}, {Finelli}, {Forni}, {Frailis}, {Fraisse},
  {Franceschi}, {Frejsel}, {Galeotta}, {Galli}, {Ganga}, {Gauthier}, {Gerbino},
  {Ghosh}, {Giard}, {Giraud-H{\'e}raud}, {Giusarma}, {Gjerl{\o}w},
  {Gonz{\'a}lez-Nuevo}, {G{\'o}rski}, {Gratton}, {Gregorio}, {Gruppuso},
  {Gudmundsson}, {Hamann}, {Hansen}, {Hanson}, {Harrison}, {Helou},
  {Henrot-Versill{\'e}}, {Hern{\'a}ndez-Monteagudo}, {Herranz}, {Hildebrand t},
  {Hivon}, {Hobson}, {Holmes}, {Hornstrup}, {Hovest}, {Huang}, {Huffenberger},
  {Hurier}, {Jaffe}, {Jaffe}, {Jones}, {Juvela}, {Keih{\"a}nen}, {Keskitalo},
  {Kisner}, {Kneissl}, {Knoche}, {Knox}, {Kunz}, {Kurki-Suonio}, {Lagache},
  {L{\"a}hteenm{\"a}ki}, {Lamarre}, {Lasenby}, {Lattanzi}, {Lawrence}, {Leahy},
  {Leonardi}, {Lesgourgues}, {Levrier}, {Lewis}, {Liguori}, {Lilje},
  {Linden-V{\o}rnle}, {L{\'o}pez-Caniego}, {Lubin}, {Mac{\'\i}as-P{\'e}rez},
  {Maggio}, {Maino}, {Mandolesi}, {Mangilli}, {Marchini}, {Maris}, {Martin},
  {Martinelli}, {Mart{\'\i}nez-Gonz{\'a}lez}, {Masi}, {Matarrese}, {McGehee},
  {Meinhold}, {Melchiorri}, {Melin}, {Mendes}, {Mennella}, {Migliaccio},
  {Millea}, {Mitra}, {Miville-Desch{\^e}nes}, {Moneti}, {Montier}, {Morgante},
  {Mortlock}, {Moss}, {Munshi}, {Murphy}, {Naselsky}, {Nati}, {Natoli},
  {Netterfield}, {N{\o}rgaard-Nielsen}, {Noviello}, {Novikov}, {Novikov},
  {Oxborrow}, {Paci}, {Pagano}, {Pajot}, {Paladini}, {Paoletti}, {Partridge},
  {Pasian}, {Patanchon}, {Pearson}, {Perdereau}, {Perotto}, {Perrotta},
  {Pettorino}, {Piacentini}, {Piat}, {Pierpaoli}, {Pietrobon}, {Plaszczynski},
  {Pointecouteau}, {Polenta}, {Popa}, {Pratt}, {Pr{\'e}zeau}, {Prunet},
  {Puget}, {Rachen}, {Reach}, {Rebolo}, {Reinecke}, {Remazeilles}, {Renault},
  {Renzi}, {Ristorcelli}, {Rocha}, {Rosset}, {Rossetti}, {Roudier},
  {Rouill{\'e} d'Orfeuil}, {Rowan-Robinson}, {Rubi{\~n}o-Mart{\'\i}n},
  {Rusholme}, {Said}, {Salvatelli}, {Salvati}, {Sandri}, {Santos},
  {Savelainen}, {Savini}, {Scott}, {Seiffert}, {Serra}, {Shellard}, {Spencer},
  {Spinelli}, {Stolyarov}, {Stompor}, {Sudiwala}, {Sunyaev}, {Sutton},
  {Suur-Uski}, {Sygnet}, {Tauber}, {Terenzi}, {Toffolatti}, {Tomasi},
  {Tristram}, {Trombetti}, {Tucci}, {Tuovinen}, {T{\"u}rler}, {Umana},
  {Valenziano}, {Valiviita}, {Van Tent}, {Vielva}, {Villa}, {Wade}, {Wandelt},
  {Wehus}, {White}, {White}, {Wilkinson}, {Yvon}, {Zacchei}, \&
  {Zonca}}]{Planck2015}
{Planck Collaboration}, {Ade}, P.~A.~R., {Aghanim}, N., {et~al.} 2016, \aap,
  594, A13

\bibitem[{Sales {et~al.}(2016)Sales, Navarro, Oman, Fattahi, Ferrero, Abadi,
  Bower, Crain, Frenk, Sawala, Schaller, Schaye, Theuns, \& White}]{Sales2016}
Sales, L.~V., Navarro, J.~F., Oman, K., {et~al.} 2016, Monthly Notices of the
  Royal Astronomical Society, 464, 2419

\bibitem[{Santos-Santos {et~al.}(2015)Santos-Santos, Brook, Stinson, Di~Cintio,
  Wadsley, Domínguez-Tenreiro, Gottlöber, \& Yepes}]{Santos-Santos2015}
Santos-Santos, I.~M., Brook, C.~B., Stinson, G., {et~al.} 2015, Monthly Notices
  of the Royal Astronomical Society, 455, 476

\bibitem[{Sawala {et~al.}(2016)Sawala, Frenk, Fattahi, Navarro, Bower, Crain,
  Vecchia, Furlong, Helly, Jenkins, Oman, Schaller, Schaye, Theuns, Trayford,
  \& White}]{Sawala2016}
Sawala, T., Frenk, C.~S., Fattahi, A., {et~al.} 2016, Monthly Notices of the
  Royal Astronomical Society, 457, 1931

\bibitem[{Schive {et~al.}(2014)Schive, Chiueh, \& Broadhurst}]{Schive2014}
Schive, H.-Y., Chiueh, T., \& Broadhurst, T. 2014, Nature Physics, 10, 496

\bibitem[{Schwabe {et~al.}(2016)Schwabe, Niemeyer, \& Engels}]{Schwabe2016}
Schwabe, B., Niemeyer, J.~C., \& Engels, J.~F. 2016, Phys. Rev. D, 94, 043513

\bibitem[{Sharma {et~al.}(2019)Sharma, Khoury, \& Lubensky}]{Sharma2019}
Sharma, A., Khoury, J., \& Lubensky, T. 2019, Journal of Cosmology and
  Astroparticle Physics, 2019, 054

\bibitem[{Skrbek(2011)}]{Skrbek2011}
Skrbek, L. 2011, Journal of Physics: Conference Series, 318, 012004

\bibitem[{Skrbek \& Sreenivasan(2012)}]{Skrbek2012}
Skrbek, L. \& Sreenivasan, K.~R. 2012, Physics of Fluids, 24, 011301

\bibitem[{Soulaine {et~al.}(2017)Soulaine, Quintard, Baudouy, \&
  Van~Weelderen}]{Soulaine2017}
Soulaine, C., Quintard, M., Baudouy, B., \& Van~Weelderen, R. 2017, Physical
  Review Letters, 118, 074506

\bibitem[{Spergel \& Steinhardt(2000)}]{Spergel2000}
Spergel, D.~N. \& Steinhardt, P.~J. 2000, Phys. Rev. Lett., 84, 3760

\bibitem[{Taylor \& Griffin(2005)}]{Taylor2005}
Taylor, E. \& Griffin, A. 2005, Phys. Rev. A, 72, 8739

\bibitem[{Tegmark {et~al.}(2004)Tegmark, Blanton, Strauss, Hoyle, Schlegel,
  Scoccimarro, Vogeley, Weinberg, Zehavi, Berlind, Budavari, Connolly,
  Eisenstein, Finkbeiner, Frieman, Gunn, Hamilton, Hui, Jain, Johnston, Kent,
  Lin, Nakajima, Nichol, Ostriker, Pope, Scranton, Seljak, Sheth, Stebbins,
  Szalay, Szapudi, Verde, Xu, Annis, Bahcall, Brinkmann, Burles, Castander,
  Csabai, Loveday, Doi, Fukugita, III, Hennessy, Hogg, Ivezi{\'{c}}, Knapp,
  Lamb, Lee, Lupton, McKay, Kunszt, Munn, O'Connell, Peoples, Pier, Richmond,
  Rockosi, Schneider, Stoughton, Tucker, Berk, Yanny, \& and}]{Tegmark2004}
Tegmark, M., Blanton, M.~R., Strauss, M.~A., {et~al.} 2004, The Astrophysical
  Journal, 606, 702

\bibitem[{Toro(2006)}]{Toro2006}
Toro, E. 2006, Applied Numerical Mathematics, 56, 1464

\bibitem[{Tulin \& Yu(2018)}]{Tulin2018}
Tulin, S. \& Yu, H.-B. 2018, Physics Reports, 730, 1 , dark matter
  self-interactions and small scale structure

\bibitem[{Zhu {et~al.}(2016)Zhu, Marinacci, Maji, Li, Springel, \&
  Hernquist}]{Zhu2016}
Zhu, Q., Marinacci, F., Maji, M., {et~al.} 2016, Monthly Notices of the Royal
  Astronomical Society, 458, 1559

\bibitem[{Zuntz {et~al.}(2010)Zuntz, Zlosnik, Bourliot, Ferreira, \&
  Starkman}]{Zuntz2010}
Zuntz, J., Zlosnik, T.~G., Bourliot, F., Ferreira, P.~G., \& Starkman, G.~D.
  2010, Phys. Rev. D, 81, 104015

\end{thebibliography}
%%%%%%%%%%%%%%%%%%%%%%%%%%%%%%%%%%%%%%%%%%%
%%%%%%%%%%%%%%%%%%%%%%%%%%%%%%%%%%%%%%%%%%%

%%%%%%%%%%%%%%%%%%%%%%%%%%%%%%%%%%%%%%%%%%%
%%%%%%%%%%%%%%%%%%%%%%%%%%%%%%%%%%%%%%%%%%%
%%%%%% Put output from bib-file here (bbl file)

%%%%%%%%%%%%%%%%%%%%%%%%%%%%%%%%%%%%%%%%%%%
%%%%%%%%%%%%%%%%%%%%%%%%%%%%%%%%%%%%%%%%%%%

%%%%%%%%%%%%%%%%%%%%%%%%%%%%%%%%%%%%%%%%%%%%%%%%%%%%%%%%%%%%%%%%%%%%%%%%%%%%%%%%%
%%%%%%%%%%%%%%%%%%%%%%%%%%%%%%%%%%%%%%%%%%%%%%%%%%%%%%%%%%%%%%%%%%%%%%%%%%%%%%%%%
%%%%%%%%%%%%%%%%%%%%%%%%%%%%%%%%%%%%%%%%%%%%%%%%%%%%%%%%%%%%%%%%%%%%%%%%%%%%%%%%%
%%%%%%%%%%%%%%%%%%%%%%%%%%%%%%%%%%%%%%%%%%%%%%%%%%%%%%%%%%%%%%%%%%%%%%%%%%%%%%%%%

\appendix
\section{equation of state}
\label{app:eos_details}

An EOS for a weakly interacting Bose gas valid at all temperatures was recently proposed by \citet{Sharma2019} for two-body and three-body interactions. Since we do not know the true EOS of DM and must resort to toy models, we instead approximated the EOS by using an ideal Bose gas with zero-temperature contributions from weak interactions. At very low temperatures, this approximation breaks down as the interactions become increasingly important, but we generally remain well above this regime.

An important quantity is the critical temperature $T_c,$ above which the fluid behaves as a normal fluid, while below the fluid condenses into a BEC and becomes superfluid;
\begin{equation}
    T_c = \frac{2\pi\hbar^2}{m k_{\text{B}}}\left(\frac{n}{\zeta(3/2)}\right)^{2/3} = \frac{2\pi\hbar^2}{m^{5/3} k_{\text{B}}}\left(\frac{\rho}{\zeta(3/2)}\right)^{2/3},
\end{equation}
where $n=\rho/m$ is the particle number density.

As an estimate for the superfluid fraction $f_s=\rho_s/\rho$ we use the fraction of particles in the BEC in an ideal Bose gas;
\begin{equation}
    f_s =
    \begin{cases}
    &1-\left(\frac{T}{T_c}\right)^{3/2}, \quad\quad T\leq T_c \\
    &0, \quad\quad\quad\quad\quad\quad T> T_c.
    \end{cases}
\end{equation}

For the other thermodynamic quantities, such as pressure, entropy, etc., we must consider them above and below $T_c$ separately. Both two-body and three-body interactions are given, parameterized by $g_2$ and $g_3$, respectively. 
\subsection{$T>T_c$}
The pressure is given by
\begin{equation}
P =
\begin{cases}
    & g_2 n^2 + \frac{\sqrt{2}\Gamma(5/2)(k_{\text{B}}T)^{5/2}m^{3/2}}{3\pi^2\hbar^3}\text{Li}_{5/2}\left(e^{\beta(\mu-2g_2 n)}\right), \,\, \text{two-body} \\
    & 4g_3 n^3 + \frac{\sqrt{2}\Gamma(5/2)(k_{\text{B}}T)^{5/2}m^{3/2}}{3\pi^2\hbar^3}\text{Li}_{5/2}\left(e^{\beta(\mu-6g_3 n^2)}\right), \,\, \text{three-body},
\end{cases}
\end{equation}
where $\Gamma(x)$ is the gamma function, $\text{Li}_{z}(x)$ is the polylogarithmic function, $\beta=1/k_{\text{B}}T$, and the chemical potential $\mu$ is determined by the equation for the number density
\begin{equation}
    n =
\begin{cases}
    & \frac{\sqrt{2}\Gamma(5/2)}{3\pi^2\hbar^3}(k_{\text{B}}T)^{3/2}m^{3/2}\text{Li}_{3/2}\left(e^{\beta(\mu-2g_2 n)}\right), \,\, \text{two-body} \\
    & \frac{\sqrt{2}\Gamma(5/2)}{3\pi^2\hbar^3}(k_{\text{B}}T)^{3/2}m^{3/2}\text{Li}_{3/2}\left(e^{\beta(\mu-6g_3 n^2)}\right), \,\, \text{three-body}.
\end{cases}
\end{equation}
The entropy is
\begin{equation}
    S = 
\begin{cases}
    & \frac{5}{2}\frac{P}{T} - n\beta(\mu-2g_2 n), \,\, \text{two-body} \\
    & \frac{5}{2}\frac{P}{T} - n\beta(\mu-6g_3 n^2), \,\, \text{three-body},
\end{cases}
\end{equation}
and the internal energy is
\begin{equation}
    U = ST - P + \mu n.
\label{eq:internal_energy}
\end{equation}

The sound speed used when determining the time-stepping in the numerical scheme was
\begin{equation}
    c_s = \sqrt{\frac{5}{3}\frac{k_{\text{B}}T}{m}}.
\end{equation}

In the limit of very high temperature, these reduce to the classical ideal gas.

\subsection{$T\leq T_c$}
The EOS below the critical temperature is given by the ideal Bose gas plus some zero-temperature contributions due to interactions;
\begin{equation}
    P = 
\begin{cases}
    & \frac{1}{2}g_2 n^2 + \zeta(5/2) \left(\frac{m}{2\pi\hbar^2}\right)^{3/2}\left(k_{\text{B}}T\right)^{5/2}, \,\, \text{two-body} \\
    & \frac{2}{3}g_3 n^3 + \zeta(5/2) \left(\frac{m}{2\pi\hbar^2}\right)^{3/2}\left(k_{\text{B}}T\right)^{5/2}, \,\, \text{three-body},
\end{cases}
\end{equation}
\begin{equation}
    S = \frac{5}{2}\zeta(5/2) \left(\frac{m}{2\pi\hbar^2}\right)^{3/2}k_{\text{B}}^{5/2}T^{3/2}, \,\, \text{two-and three-body},
\end{equation}
\begin{equation}
    \mu =
\begin{cases}
    & g_2 n, \,\, \text{two-body} \\
    & g_3 n^2, \,\, \text{three-body},
\end{cases}
\end{equation}
and the internal energy is again given by Eq. \eqref{eq:internal_energy}. The fastest sound speed was approximated using
\begin{equation}
    c_s = \sqrt{\frac{\zeta(5/2)}{\zeta(3/2)}\frac{5}{3}\frac{k_{\text{B}}T}{m}},
\end{equation}
and the critical velocity given by
\begin{equation}
    v_c = 
\begin{cases}
    & \sqrt{\frac{g_2 n}{m}\left[1-(T/T_c)^{3/2}\right]}, \,\, \text{two-body} \\
    & \sqrt{\frac{2g_3 n^2}{m}\left[1-(T/T_c)^{3/2}\right]\left[1+2(T/T_c)^{3/2}\right]}, \,\, \text{three-body.}
\end{cases}
\end{equation}

There is a small discontinuity at the critical temperature, with $\mu=2ng_2$ above and $\mu=ng_2$ below for the two-body interaction (and a similar jump in zero-temperature pressure and internal energy). There should be a crossover region as the condensate fraction increases, but during this crossover the thermal contributions dominates and the discontinuity is negligible.

\section{numerical scheme}
\label{app:numerical_scheme}

In this work, we employed a modified first-order FORCE scheme \citep{Toro2006}---an incomplete Riemann solver---for the superfluid hydrodynamic equations with source terms due to gravity and from using spherical coordinates. The source terms were evaluated at two stages during each time-step: once before the advection step, and once after, at which point the average of the two evaluations was added to the solution. Gravity was also evaluated with half a time-step when computing fluxes during the advection step. Finally, we enforced the critical velocity, which was done in three stages; once when computing fluxes, once after the fluxes from the advection step were applied, and a final time after the source terms were applied. Further details are presented below.

For spherical collapse, this scheme was found to be sufficient since the solutions are mostly smooth, evolve slowly, and are one-dimensional. For more complex and higher dimensional cases where shock fronts arise and the solutions undergo fast changes, this scheme is expected to perform suboptimally, primarily because it is first-order. There is a well-known way to increase the order and thus accuracy of the scheme through slope reconstruction and slope limiters. However, instabilities arose when the superfluid component was included, and adding further restrictions to the reconstructed slopes with modified slope limiters failed to fix this. Slope reconstruction was therefore not used.

\subsection{first-order FORCE scheme}

The FORCE scheme is a variant of Godunov's method for solving partial differential equations. In this method, the domain is divided into finite-volume elements, or cells, and the Riemann problem at each cell interface is solved. The Riemann problem is the initial value problem with two piece-wise constant initial regions connected by a discontinuity, then asking how this evolves in time and what the net flux across the interface is. The scheme for computing or approximating this flux is called a Riemann solver and is what characterizes the different ways of implementing Godunov's method.

To see how this works, one can consider the m-component state vector U that obeys the one-dimensional conservative equation
\begin{equation}
    \partial_t \bm{U} + \partial_x \bm{F}(\bm{U}) = \bm{0},
\end{equation}
where $\bm{F}$ is the flux. By integrating over the time interval $[t^n, t^{n+1}]$ and cell-volume $[x_{i-1/2}, x_{i+1/2}],$ we get
\begin{equation}
    \bm{U}^{n+1}_{i} = \bm{U}^{n}_{i} - \frac{\Delta t}{\Delta x}[\bm{F}_{i-1/2} - \bm{F}_{i+1/2}],
\end{equation}
where 
\begin{equation}
    \bm{U}^{n}_{i} = \frac{1}{\Delta x}\int_{x_{i-1/2}}^{x_{i+1/2}} \bm{U}(x,t^n)\text{d}x,
\end{equation}
\begin{equation}
    \bm{F}_{i+1/2} = \frac{1}{\Delta t}\int_{t^{n}}^{t^{n+1}} \bm{F}(x_{i+1/2},t)\text{d}t.
\end{equation}
In the first-order Godunov scheme, the state $\bm{U}$ is assumed to be piece-wise constant in each cell, given by the cell average $\bm{U}^{n}_{i}$. To compute $\bm{F}_{i+1/2,}$ the states on the left and right sides of the interface is used, $\bm{U}_{i+1/2,L} = \bm{U}^n_{i}$ and $\bm{U}_{i+1/2,R} = \bm{U}^n_{i+1}$, and the corresponding Riemann problem is solved or approximated. The time-step is chosen so that no signal in the domain travels further than one cell length $\Delta x$. This is given by a Courant-Friedrich-Lewy (CFL) type condition
\begin{equation}
    \Delta t_{\text{s}} = C^{\text{s}}_{\text{CFL}} \frac{\Delta x}{v_{\text{max}}},
\end{equation}
where $v_{\text{max}}$ is the maximum signal speed in the domain, and $C^{\text{s}}_{\text{CFL}}$ is a number less than one that controls how far across a cell the fastest signal is allowed to move during each time-step. In simulations with gravity and expansion, additional constraints need to be added to the time-stepping. For gravity, the free-fall distance in each cell, with acceleration $g$, must be smaller than the cell lengths,
\begin{equation}
    \Delta t_{\text{ff}} = C^{\text{ff}}_{\text{CFL}}\sqrt{\frac{2\Delta x}{g}},
\end{equation}
and for expansion, the relative change in the scale factor is restricted:
\begin{equation}
    \Delta t_{\text{exp}} = C^{\text{exp}}_{\text{CFL}} \frac{1}{H}.
\end{equation}
Here, $C^{\text{ff}}_{\text{CFL}}$ and $C^{\text{exp}}_{\text{CFL}}$ are also numbers less than one. In this work, we used $C^{\text{s}}_{\text{CFL}}=0.5$, $C^{\text{ff}}_{\text{CFL}} = 0.5$, and $C^{\text{exp}}_{\text{CFL}}=0.01$.
The final value for the time-step is the smallest of the above,
 \begin{equation}
     \Delta t = \min[ \Delta t_{\text{s}},\, \Delta t_{\text{ff}},\, \Delta t_{\text{exp}}]
 .\end{equation}
 
The FORCE scheme approximates the interface flux $\bm{F}$ (given the left and right states $\bm{U}_L$ and $\bm{U}_R$) as the average of the Lax-Friedrichs flux and the two-step Lax-Wendroff flux;
\begin{equation}
\begin{split}
    &\bm{F}^{\text{FORCE}} = \frac{1}{2}[\bm{F}^{\text{LF}} + \bm{F}^{\text{LW}}], \\
    &\bm{F}^{\text{LF}} = \frac{1}{2}[\bm{F}(\bm{U}_{L}) + \bm{F}(\bm{U}_{R})] - \frac{1}{2}\frac{\Delta x}{\Delta t}[\bm{U}_{R} - \bm{U}_{L}], \\
    &\bm{F}^{\text{LW}} = \bm{F}(\bm{U}^{\text{LW}}),\\
    &\bm{U}^{\text{LW}} = \frac{1}{2}[\bm{U}_{L}  + \bm{U}_{R}] - \frac{1}{2}\frac{\Delta t}{\Delta x}[\bm{F}(\bm{U}_{R}) - \bm{F}(\bm{U}_{L})].
\end{split}
\end{equation}
We modified this by enforcing the critical velocity on the intermediate state $\bm{U}^{\text{LW}}$ before computing the flux $\bm{F}^{\text{LW}}$.

\subsection{sources}
Gravity and extra terms when using spherical coordinates and super-comoving variables appear as source terms $\bm{S}$ in the superfluid equations. Continuing with the above example, we have
\begin{equation}
    \partial_t \bm{U} + \partial_x \bm{F}(\bm{U}) = \bm{S}.
\end{equation}
To modify our Godunov scheme to incorporate the sources in the flux, we did the following: at the beginning of each time-step, we had the states $\bm{U}^n_i$. To do the advection (the Godunov step), we input the left and right states at each boundary $i+1/2$; $\bm{U}_{i+1/2,L} = \bm{U}_{i}$, $\bm{U}_{i+1/2,R} = \bm{U}_{i+1}$. But before we computed the interface flux, we applied half a time-step of the source due to gravity,
\begin{equation}
\begin{split}
    &\bm{U}^{*}_{i+1/2, L} = \bm{U}_{i+1/2, L} + \frac{1}{2}\Delta t \left(\bm{S}^n_{i+1/2, L}\right)_{\text{grav}},\\ &\bm{U}^{*}_{i+1/2, L} = \bm{U}_{i+1/2, R} + \frac{1}{2}\Delta t \left(\bm{S}^n_{i+1/2, R}\right)_{\text{grav}},
\end{split}
\end{equation}
where $\bm{S}^n_{i+1/2, L}$ and $\bm{S}^n_{i+1/2, R}$ are the left and right values for the sources. In this work, we computed these using the average gravitational acceleration $(\bm{\nabla}\Phi)^{n}_{i+1/2,L/R} = \frac{1}{2}[(\bm{\nabla}\Phi)^{n}_{i} + (\bm{\nabla}\Phi)^{n}_{i+1}]$, and the left and right states $\bm{U}_{i+1/2, L/R}$.
We then used $\bm{U}^{*}_{i+1/2, L}$ and $\bm{U}^{*}_{i+1/2, R}$ as the input states in the Godunov scheme to get $\bm{F}_{i+1/2}$, and updated the state vectors from the previous time-step:
\begin{equation}
    \bm{U}^{n+1,*}_{i} = \bm{U}^{n}_{i} - \frac{\Delta t}{\Delta x}[\bm{F}_{i-1/2} - \bm{F}_{i+1/2}].
\end{equation}

This modification to the Godunov scheme was to include the effect of gravity on the flux, but explicitly adding the sources to the solution remains to be done. For this, we used the average before and after the advection step;
\begin{equation}
    \bm{U}^{n+1}_{i} = \bm{U}^{n+1,*}_{i} + \Delta t \frac{1}{2} [\bm{S}^{n}_{i} + \bm{S}^{n+1,*}_{i}].
\end{equation}

\subsection{enforcing critical velocity}
\label{app:numerical_enforce_vc}
The critical velocity was enforced by iteratively converting kinetic energy into internal energy and generated entropy in all cells until $w<v_c$. The scheme works as follows: we consider a cell with the state vector $\bm{U}^l$, where $l$ denotes the current step in the iterative scheme to enforce $v_c$, and $l=0$ is the initial state. From this we get the fluid variables of the cell, $v_s^l$, $v_n^l$, $S^l$, etc. If $w^l < v_c^l$, the Landau criterion is satisfied and we do nothing. If instead $w^l > v_c^l$, we apply a small change $\Delta v_s^l$ to $v_s^l$ to update it to $l+1$ and decrease $w$,
\begin{equation}
    \bm{v}^{l+1}_{s} = \bm{v}^l_{s} + \Delta \bm{v}_s^l.
\end{equation}
By keeping $\bm{j}$ constant and assuming that the change in the superfluid fraction is negligible compared to the change in velocity, we get
\begin{equation}
    \Delta \bm{v}_n^l = -\frac{\rho_s^l}{\rho_n^l}\Delta\bm{v}_s^l.
\end{equation}
Using conservation of energy the change in internal energy is equal to the change in kinetic energy;
\begin{equation}
\begin{split}
    \Delta U^l &= -\Delta E_{\text{kin}}^l = -\Delta\Bigg(\frac{1}{2}\rho_n^l |\bm{v}_n^l|^2 + \frac{1}{2}\rho_s^l |\bm{v}_s^l|^2\Bigg)\\
    &=-\rho_s^l(\bm{v}_s^l - \bm{v}_n^l) \cdot \Delta \bm{v}_s^l.
\end{split}
\end{equation}
The change in entropy is $\Delta S^l = \Delta Q^l / T^l$, where $\Delta Q^l$ is the heating of the fluid, which in this case is just the change in internal energy:
\begin{equation}
    \Delta S^l = \frac{\Delta Q^l}{T^l} = \frac{\Delta U^l}{T^l} = -\frac{\rho_s^l(\bm{v}_s^l - \bm{v}_n^l) \cdot \Delta \bm{v}_s^l}{T^l}.
\end{equation}
The updated entropy is
\begin{equation}
    S^{l+1} = S^l + \Delta S^l.
\end{equation}
We arrive at the state vector $\bm{U}^{l+1}$ and repeat the above process until $w < v_c$. The only part that needs to be specified is $\Delta \bm{v}_s^l$, which was chosen as
\begin{equation}
    \Delta \bm{v}_s^l = 
    -C(w^l)\,\hat{\bm{w}}^l,
\end{equation}
where
\begin{equation}
    C(w) = [10^{-2}, 10^{-5}] w.
    \label{eq:delta_vs_formula}
\end{equation}
The numerical factor in Eq. (\ref{eq:delta_vs_formula}) was tuned to give as smooth $w$-profile as possible while keeping the scheme from becoming too slow.

%\end{appendix}

%%%% End of aa.dem
\end{document}